\documentclass{elsarticle}

\pdfoutput=1

\usepackage{amssymb}
\usepackage{amsthm}
\usepackage{graphicx}
\usepackage{epsfig}
\usepackage{latexsym, amsmath}
\usepackage[english]{babel}
\usepackage{amsmath, amstext, amsfonts}
\usepackage{float}
\usepackage{caption}
\usepackage{subcaption}
\usepackage{appendix}
\usepackage{xcolor}

\def\R{{\mathbf R}}

\newcommand{\x}{\mathbf{x}}

\newtheorem{theorem}{Theorem}[section]

\newtheorem{definition}[theorem]{Definition}

\journal{ .}

\begin{document}

\begin{frontmatter}

\title{
Electronic implementation of a dynamical network with nearly identical hybrid nodes via unstable dissipative systems}

\author[buap]{A.~Anzo-Hern\'andez}
\ead{andres.anzo@hotmail.com}

\author[itesm]{M. Garc\'ia-Mart\'inez}
\ead{moises.garcia@itesm.mx}

\author[ipicyt]{E.~Campos-Cant\'on }
\ead{eric.campos@ipicyt.edu.mx}

\author[altiplano]{L.J.~Onta\~n\'on-Garc\'ia\corref{cor1}}
\ead{luis.ontanon@uaslp.mx}
\cortext[cor1]{Corresponding author}

\address[buap]{
{\textsc{C\'atedras CONACYT - Benem\'erita Universidad Aut\'onoma de Puebla - Facultad de Ciencias F\'isico-Matem\'aticas},\\}
{Benemerita Universidad Aut\'onoma de Puebla},\\
\textsc{Avenida San Claudio y 18 Sur, Colonia San Manuel, 72570.\\
Puebla, Puebla, M\'exico.
\vskip 2ex}}

\address[itesm]{
 \textsc{Tecnologico de Monterrey,\\}
 {Escuela de ingenier\'ia y ciencias}\\
 San Luis Potos\'{\i}, SLP, M\'{e}xico \vskip 2ex}

\address[ipicyt]{
 \textsc{IPICYT/Divisi\'on de Matem\'aticas Aplicadas ,\\}
  \textsc{Camino a la Presa San Jos\'e 2055 col. Lomas 4a Secci\'on, 78216,\\
 San Luis Potos\'{\i}, SLP, M\'{e}xico \vskip 2ex}}

\address[altiplano]{
{\textsc{Coordinaci\'on Acad\'emica Regi\'on Altiplano Oeste},\\}
{Universidad Aut\'onoma de San Luis Potos\'i},\\
\textsc{Kilometro 1 Carretera a Santo Domingo, 78600\\
Salinas de Hidalgo, San Luis Potos\'i , M\'exico.
\vskip 2ex}}

\setcounter{page}{1}

\begin{abstract}
A circuit architecture is proposed and implemented for a dynamical network composed of a type of hybrid chaotic oscillator based on Unstable Dissipative Systems (UDS).  The circuit architecture allows selecting a network topology with its link attributes and to study, experimentally, the practical synchronous collective behavior phenomena. Additionally, based on the theory of dynamical networks, a mathematical model of the circuit was described, taking  into account the natural tolerance of the electronic components.
The network is analyzed both numerically and experimentally according to the parameters mismatch between nodes.
\end{abstract}


\end{frontmatter}

\section{Introduction}

The design and circuit implementation of chaotic oscillators have been an active research subject that allows  to better understand the chaos phenomena and influencing it.  Several technological applications of such chaotic circuits have been also proposed in the area of  secure communications or in  cryptography devices \cite{Cuomo1993,Strogatz1993,Garcia-Martinez2015}.  Another advantages of electronic circuits is its feasibility to represent chaotic dynamics throughout analog circuitry. Take for instance the studies published in \cite{Orponen1997,Siegelmann1998,Horio2008}, where the electronic circuit implementation of a dynamical system is simplified by means of Operational Amplifiers (Op Amp's).

Other form to reduce the circuit design complexity is to use linear dynamical systems and induce a chaotic behavior by the introduction of simple switching law. Such systems are usually called Piece-Wise Linear (PWL) systems and are capable of generating various scroll attractors in the phase space. One of the most prominent example of PWL system is the so called Chua's circuit \cite{Mulukutla2002} which has also stimulated the current research interest in creating numerous chaotic circuit with simple electronic components. Other example is the PWL system based on unstable dissipative system (UDS) which is constructed from the jerk equations and its simple mathematical expression allows {its implementation with few electronic} components \cite{Campos-Canton2010}.

In recent years, some research efforts have been focused on implementing circuit devices that emulate the dynamics of two or more coupled chaotic circuits. For example, Mu\~noz-Pacheco {\it et al}  \cite{Munoz-Pacheco2014}  used metal-oxide-semiconductor (CMOS) integrated circuit technology in order to fabricate multi-scroll oscillators coupled in a master-slave configuration. The field-programmable gate arrays (FPGAs) have been also used to realize multi-scroll chaotic oscillators in \cite{Tlelo-Cuautle2015}. In the same direction, Leyva \textit{et. al.} \cite{Leyva2012} implemented a network of electronic R\"ossler oscillators coupled in a star configuration  and Magistris \textit{et. al.} \cite{DeMagistris2012a} investigated the robustness of an ensemble of coupled non-identical Chua's circuits.

The coupled chaotic circuits studied and implemented in the aforementioned research works, form part of what is usually called dynamical complex networks (or simply dynamical networks).  The term complexity is introduced here with the aim to highlight that these systems are composed by dynamical units coupled in a non trivial topology configuration \cite{Strogatz2001}. One of the collective behavior that emerges in dynamical  complex network is synchronization which occurs when the motion of all the individuals in the net follows a common evolution \cite{Boccaletti2006a,Porter2014}. Some {criteria} which are often used to detect  complete synchronization are the second-largest eigenvalue condition of the coupling matrix \cite{Wang2003} or the Master Stability Function \cite{Pecora1998}. However,  these methods have been derived for networks composed of identically chaotic oscillators, so the structure of the network plays a crucial role in determining if the synchronization occurs or not.

Notwithstanding the vast literature on the synchronization of dynamical networks, the majority of the existing studies usually assume that nodes are identical, that is, the dynamics of each node is described by the same vector field. However, {taking into account the diversity that nature presents, it is often a difficult task or attributed to the luck of finding} two systems with exactly the same characteristics. In light of this, some reports have been made on the study of non-identical nodes among networks \cite{DeMagistris2012a,Zhao2011}. This comes as a major concern in different areas such as biology and sociology, but in the particular case of electronics, it is well known that the manufacturer industry fabricates passive components such as  resistors, capacitors, etc.,  that present specific tolerances in their nominal values, {{\it i. e.}}, the value of each component {differs}  from the value reported by the manufacturer. In this scenario, a resistor with a nominal value of $100 \Omega$ and a tolerance of $10\%$ will actually have a value between $90 \Omega$ and $110 \Omega$, as have been pointed out in \cite{RobertSpence1997}.

Considering the problem described above, if an electronic circuit is designed in order to emulate a dynamical complex network composed of identically chaotic oscillators, then their corresponding nodes parameters will exhibit variations that make them not identical at all. It is worth to mention that a given chaotic system is sensitive, amongst other things, to small changes in their parameters values. Such situations motivated the present study in the relationship among the nodes with parameter mismatch in a dynamical network in order to detect the emergence of synchronization. Each node of the network will be considered as a PWL system based on  UDS following two methodologies to study the synchronization: 1) based on the formalism of dynamical complex networks, a mathematical model where  each node present variations on their parameter configuration due to the intrinsic tolerances on their components will be proposed, resulting in a nearly identical network; 2) an electronic circuit using Op Amps will be designed and implemented in order to study the physical nearly identical network in a more tangible manner. The configuration of the network connections can be varied among the nodes by means of physical connections in the circuits using wires and resistances. In particular, two network topologies were tested, namely Fully Connected (FC) and Nearest Neighbor (NN) configurations. Additionally, from each node a randomly parameter variation of $\pm 20 \%$ was introduced from a nominal value. The numerical and electronic circuit experiment have shown that despite their parametric mismatch, the network achieve a bounded practical synchronization, that is, the differences between nodes variables states are less that an small positive number.

This paper is organized as follows: in Section 2 is described the main attributes of a hybrid system based on UDS. In Section 3, some basic preliminaries about dynamical complex networks and synchronization phenomena are introduced. Next, Section 4 present the problem statement where the model of a network of nearly identical hybrid nodes is described. In Section 5, the numerical results and in Section 6 the design of the circuit and the corresponding results of its performances are shown. Finally, in Section 7 some concluding remarks are discussed.

\section{Hybrid dynamical system}

In order to define a hybrid system based on unstable dissipative systems (UDS) in a similar way as \cite{UDS4},  the following hybrid dynamical system will be considered:
\begin{equation}\label{eq:PWL}
\dot{\x} = \mathbf{A}\x + \mathbf{B}(\x);
\end{equation}

\noindent where $\x= [x_{1},x_{2},x_ {3}]^{\top} \in \R^{3}$ is the state vector,  the matrix $\mathbf{A}  = \mathcal{M}_{3\times 3}(\{ a_{ij} \})$  is the linear operator whose entries are defined by the parameters $a_{ij} \in \R$ (for $i,j=1,2,3$); and $\mathbf{B}(\cdot): \mathbf{R}^{3} \rightarrow \mathbf{R}^{3}$ is a piecewise constant vector which is determined by a discrete dynamics behaviour of the state vector in different domains.
In order to define it, the state space $\R^{3}$ will be divided into a finite number of domains $\mathcal{S}_{k}\subset \R^3$ in a way that $\R^3=\cup_{k=1}^m\mathcal{S}_{k}$ and $\cap_{k=1}^m\mathcal{S}_{k}=\emptyset$. Therefore, the affine vector $\mathbf{\mathbf{B}}$ must be considered as a discrete function that changes depending on which domain $\mathcal{S}_k$, $k=1,\ldots,m$, the trajectory $\phi^t(\x_0)$ is located. $\phi^t$ is the flow of the system \eqref{eq:PWL} and $\x_0$ is the initial condition.
In particular, it is assumed that the system \eqref{eq:PWL} is based on the jerk equation (see \cite{Campos-Canton2012} and \cite{Njitacke2016})
$\dddot{\x} + a_{33}\ddot{\x} + a_{32}\dot{\x} + a_{31}\x + f(\x) = 0$, which can be expressed as a system of first order differential equations in the form of \eqref{eq:PWL} considering:
\begin{equation}\label{eq:uds}
   \mathbf{A} = \left( \begin{array}{ccc}
          0 & 1 & 0 \\
          0 & 0 & 1 \\
         -a_{31} & -a_{32} & -a_{33} \end{array} \right), \quad
  \mathbf{B}(\x) =  \left( \begin{array}{c}
           0 \\
           0 \\
           b_3(\x)
          \end{array} \right);
\end{equation}

\noindent where $b_3(\x): \mathbf{R}^{3} \rightarrow \mathbf{R}$ is a piecewise-constant function that controls the discrete transitions of the affine vector $\mathbf{\mathbf{B}}$ called the switching law.
With the  state space  partition $\mathcal{S}_k$, with $k=1,\ldots,m$, the switching law takes the form:
\begin{equation}\label{eq:switching_law}
b_3(\x) =  \left\{
\begin{array}{lll}
  \beta_{1}, & \text{ if }    & \x \in S_{1} = \{\x \in \mathbf{R}^{3}:   \mathbf{v}^{\top}\x < \delta_{1} \}; \\
   \beta_{2}, & \text{ if }    & \x \in S_{2}  = \{\x \in \mathbf{R}^{3}:  \delta_{1}  \leq \mathbf{v}^{\top}\x < \delta_{2} \}; \\
     \vdots    &       &  \vdots \\
   \beta_{m}, & \text{ if }  & \x \in S_{m} = \{\x \in \mathbf{R}^{3}:  \delta_{m-1}  \leq \mathbf{v}^{\top} \x \};\\
\end{array} \right.
\end{equation}


\noindent with $\beta_{k} \in \mathbf{R}$, for $k = 1,\ldots,m$. The domains in which the space is partitioned are given by $S_{i}$, where $\mathbf{v}= [v_1,v_2,v_3]^{\top}\in \mathbf{R}^{3}$ (with $\mathbf{v} \neq 0$) is a constant vector and $\delta_1 \leq \delta_2 \leq \cdots \leq \delta_{m-1} $ are scalars that define the switching regions. The switching surfaces are given by the hyperplanes $\mathbf{v}^{\top} \x = \delta_{i}$ (for $i = 1,2,\ldots,m-1$), {\it e.g.}, $v_1\cdot x_1+v_2\cdot x_2+v_3\cdot x_3-\delta_{i}=0$. For simplicity and for this particular type of systems the hyperplanes will be adjusted with $\mathbf{v} = [1,0,0]^{\top} \in \mathbf{R}^{3}$, locating them only along $x_1$ axis.

\begin{figure}[!t]
\centering
\hspace{-25pt}\includegraphics[width=12cm,height=8cm]{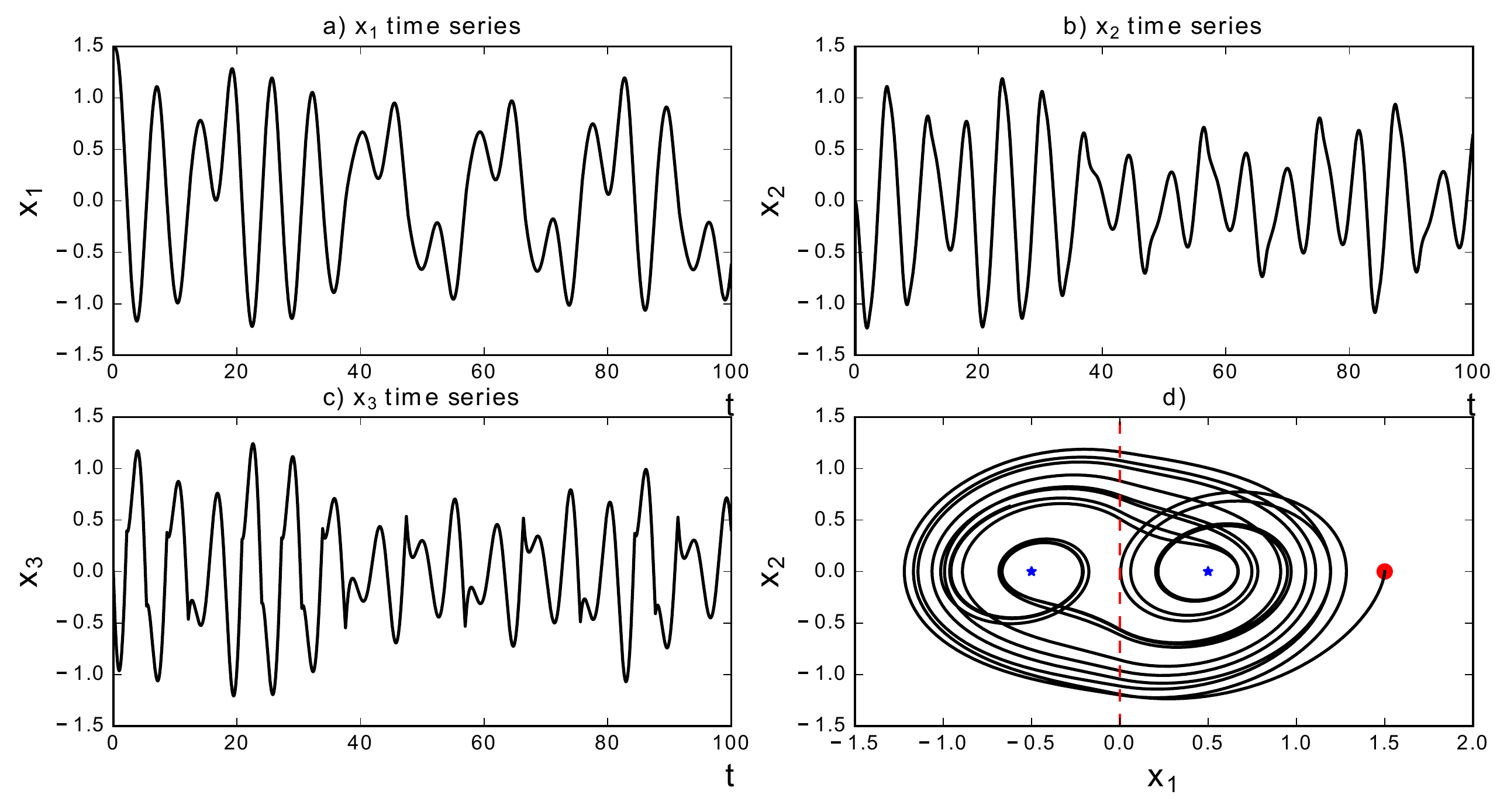}
\caption{Time series of the states: a) $x_1$, b) $x_2$, c) $x_3$. d) Projection of the trajectory onto the $(x_{1},x_{2})$ plane. The parameter of the linear operator are  $a_{31} = -2$, $a_{32} = -1$, $a_{33} = -1$ and switching law \eqref{eq:switching_law_2} . The red dashed line in Figure d) corresponds to the switching surface, the blue stars indicate the location of the equilibria  and, the initial condition at $\x_{0} = (1.5, 0, 0)^{\top}$ is marked with a red dot.}
\label{fig:Fig1}
\end{figure}

The role of the switching law is to control the discrete transition of $\mathbf{\mathbf{B}}$ in order to indicate the affine linear system \eqref{eq:PWL} that  is active, {\it i.e.}, if $b_3(\x) = \beta_{k}$ for $k \in I = \{1,\ldots,m\}$ and $\x \in S_{k}$, then the affine linear system that governs the dynamics in the $k$-th domain $S_k$  is: $\dot{\x} = \mathbf{A}\x + [0,0,\beta_{k}]^{\top}$.

It is worth to note that the system \eqref{eq:PWL}-\eqref{eq:uds} contains for each domain $S_{k}$, a single equilibrium point located at $\mathbf{\x}_{k}^{*} = -\mathbf{A}^{-1}\mathbf{B}_{k} = [\beta_{k}/a_{31},0,0]^{\top}$. In particular, the interest of the work is that each domain will be a hyperbolic set containing a single unstable focus-saddle equilibrium point $\mathbf{\x}^*$ which presents a stable manifold $M^s = span\{\varepsilon_1\}\in \R^3$ with a fast eigendirection and an unstable manifold $M^u= span\{\varepsilon_2,\varepsilon_3\}\in \R^3$ with a slow spiral eigendirection.  In this form it is ensured that for any initial condition $\x_{0} \in \mathbf{R}^{n}$, the orbit $\phi^t(\x_{0})$ of the system \eqref{eq:PWL}-\eqref{eq:uds}  is confined in an one-spiral trajectory in the region $S_{k}$ called scroll  until its size increases due to the unstable eigenspectra. When the trajectory $\phi^t(\x_0) \subset S_\rho$ reaches to the  hyperplane $x_{1} = \delta_{\rho}$, it  crosses to the region $S_{\rho+1}$, where it is again confined in a new scroll with equilibrium point  $\x_{\rho+1}^{*} = -\mathbf{A}^{-1}\mathbf{B}_{\rho+1}$ until the trajectory expands again. In this context, the system \eqref{eq:PWL}-\eqref{eq:uds} can display various multi-scroll attractors as a result of a combination of several unstable one-spiral trajectories, where the switching between regions is governed by the switching function \eqref{eq:switching_law}. In order to guarantee the existence of saddle equilibrium points, the following assumptions about the hybrid dynamical system \eqref{eq:PWL} must be considered:

\textbf{Assumptions 1}: The eigenvalues of the linear operator $\mathbf{A}$,  denoted by $\lambda_{i}$ (for $i=1,2,3$), of a linear system $\dot{x}=Ax$ satisfies:
\begin{enumerate}
   \item[A1.-] {One} of its eigenvalues (labeled as $\lambda_{1}$) is a real number;
   \item[A2.-] {The other two} of its eigenvalues (labeled as $\lambda_{2}$ and $\lambda_{3}$) are complex conjugate  and;
   \item[A3.-] The sum of its eigenvalues satisfy: $\sum^{3}_{i=1}  \lambda_{i}  < 0$.
\end{enumerate}

A hybrid dynamical system of the form \eqref{eq:PWL} that satisfy the requirements of \textbf{Assumption 1} is called a hybrid dynamical system based on unstable dissipative system (UDS). According to eigenvalues attributes of $\mathbf{A}$, it has been proposed the following classifications from UDS \cite{Campos-Canton2010}:
\begin{definition}\label{Definition:1}
The system \eqref{eq:PWL} is based on UDS \textit{Type I} if the eigenvalues of its linear operator $\mathbf{A}$ satisfy the \textbf{Assumptions 1} with $\lambda_{1}<0$, $\textit{Re} \{ \lambda_{2} \} > 0$ and $\textit{Re} \{ \lambda_{3} \} > 0$.
\end{definition}

\begin{definition}\label{Definition:2}
The system \eqref{eq:PWL} is based on UDS \textit{Type II} if the eigenvalues of its linear operator $\mathbf{A}$ satisfy the \textbf{Assumptions 1} with $\lambda_{1}>0$, $\textit{Re} \{ \lambda_{2} \} < 0$ and $\textit{Re} \{ \lambda_{3} \} < 0$.
\end{definition}


As the authors in \cite{Jimenez-Lopez2013} mentioned, the above Definitions \ref{Definition:1} and \ref{Definition:2} imply that the UDS \textit{Type I} is stable in one of its components but unstable in the other two, which are oscillatory. The converse is the UDS \textit{Type II}, which are stable and oscillatory in two of its components but unstable in the other one.

In order to illustrate this, the  dynamical system \eqref{eq:PWL}-\eqref{eq:uds} will be considered throughout the work only satisfying {Definition \ref{Definition:1}} with parameters $a_{31} = 2$, $a_{32} = 1$ and  $a_{33} = 1$ for the linear operator $\mathbf{A}$ and a switching law as follows:
\begin{equation}\label{eq:switching_law_2}
b_3(\x) =  \left\{
\begin{array}{rll}
   1,  & \text{if}  & \x \in S_{1} = \{\x \in \mathbf{R}^{3}: x_{1} \geq 0\}; \\
  -1,  & \text{if}  & \x \in S_{2} = \{\x \in \mathbf{R}^{3}: x_{1} <      0\}.  \\
\end{array} \right .
\end{equation}

For {these} parameter values, the linear operator has the following eigenvalues:  $\lambda_{1} = -1.353$ and $\lambda_{2,3} = 0.176  \pm 1.2 \textit{i}$, which according to Definition \ref{Definition:1}, the system is an UDS of \textit{Type I}. In Figure \ref{fig:Fig1} it is depicted the time series for each variable state and the projection onto the plane ($x_{1},x_{2}$) of the trajectory of the hybrid dynamical system via UDS with eqs. \eqref{eq:PWL}-\eqref{eq:uds} and the switching law \eqref{eq:switching_law_2} with initial condition $\x_{0} = (1.5,0, 0)^{\top}$. It is worth to note that the system generates a double-scroll attractor around the equilibria  $\x^{*}_{1} =  (0.5,0,0)^\top$ and $\x^{*}_{2} =  (-0.5,0,0)^\top$ due to the aforementioned switching dynamics.


\section{Dynamical networks and synchronization}

In general, a dynamical network is composed by a set of N-coupled dynamical systems of the form $\dot{\x}_{i}(t) = f_i(\x_{i}(t),\alpha_i)$; where $\x_i(t) = [x_{i1},x_{i2},\ldots,x_{in}]^{T} \in \mathbf{R}^{n}$ is the state vector of the $i$-th node; $f_{i}: \mathbf{R}^{n} \rightarrow \mathbf{R}^{n}$ is the vector field for the $i$-th dimensional isolated node; and $\alpha_{i} \in \R$ is the corresponding
mismatch parameter of $i$-th node. The parameters $\alpha_i$, with $i=1,\cdots,N$, set the difference between nearly identical hybrid nodes. In particular, it is usual to consider the case in which the nodes are identical, \textit{i.e.}, $f(\cdot) = f_{i}(\cdot)$ and $\alpha_{i}=\alpha_{j}$ $\forall i,j$. Additionally, the coupling between neighboring nodes is assumed to be bidirectional links such that the state equation of the entire network is:
\begin{equation}\label{eq:dyNet}
\dot{\x}_{i}(t) = f(\x_{i}(t), \alpha_i) + c_i \sum_{j = 1}^{N} \Delta_{ij} \Gamma (\x_{j}(t) - \x_{i}(t)), \quad i = 1,\ldots,N;
\end{equation}

\noindent where  $c_i$ is the  coupling strength; $\Gamma = diag\{r_1,\ldots,r_n \} \in \mathbf{R}^{n \times n}$ is the inner coupling matrix where $r_{l} = 1$ if nodes are linked through their $l$-th state variable, and $r_{l} = 0$ otherwise.
The network topology shows how the nodes are connected to each other, and it is described mathematically by the coupling matrix $\Delta = \mathcal{M}_{N\times N}(\{ \Delta_{ij} \})$, whose elements are zero or one depending on which pair of nodes are connected or not. This matrix contains the information of the entire network's topology and it is constructed as follows: because we consider bidirectional couplings, if there is a connection between node $i$ and node $j$ (with $i \neq j$), then $\Delta_{ij} = \Delta_{ji} = 1$; otherwise $\Delta_{ij} = \Delta_{ji} = 0$. To complete the construction of $\Delta$, their diagonal entries are calculated as
\begin{equation}\label{eq:diffusive}
\Delta_{ii} = -\sum^{N}_{\substack{j=1 \\ j \neq i}} \Delta_{ij} = -\sum^{N}_{\substack{j=1 \\ j \neq i}} \Delta_{ji}, \quad i = 1,2,\ldots,N.
\end{equation}

Eq. (\ref{eq:diffusive}) is known as the diffusive condition.  If there are not isolated nodes in the network, then $\Delta$ is a symmetric and irreducible matrix whose eigenspectrum satisfies the following conditions \cite{Wang2003}:  zero is an eigenvalue of multiplicity one; all its non zero eigenvalues are strictly negative and; they can be ordered as $0 = \mu_{1} > \mu_{2} \geq \mu_{3} \geq \cdots \geq \mu_{N}$.

For the dynamical network \eqref{eq:dyNet}, complete synchronization emerges when the state variables of each node evolve at unison in  common trajectories, and the following limit is satisfied  $\lim_{t \rightarrow \infty} || \x_{i}(t) - \x_{j}(t) || = 0$, $\forall i,j$, where $|| \cdot ||$ denotes the Euclidean norm in $\mathbf{R}^{3}$. In this case, the node's trajectory  converge asymptotically towards the synchronization manifold $\Omega = \{ \x_{i} \in \R^{3}: \x_{1} = \x_{2} =  \cdots = \x_{N} \}$. However, such type of synchronizations occurs when the nodes are identically \cite{DeMagistris2012a}. For non identical (or almost identically) nodes,  the following definition is used:

\begin{definition}\label{Definition:3}
The dynamical network \eqref{eq:dyNet} is said to achieve {practical synchronous collective behavior}  if any of the trajectories of the system nodes satisfy the following condition:
\begin{equation}\label{eq:syncro_def}
  \lim_{t \rightarrow \infty} || e_i(c)  || \leq \epsilon  \quad i = 1,\ldots,N
\end{equation}

\noindent where $e_i(c)$ correspond to the error given by the {Euclidean} distance for a given coupling strength
\begin{equation}\label{eq:syncro_def2}
e_i(c)= \sqrt{ (\x_{i1}(t) - \x_{mean_1}(t))^2 + (\x_{i2}(t) - \x_{mean_2}(t))^2 + (\x_{i3}(t) - \x_{mean_3}(t))^2},
\end{equation}

\noindent  and {$\x_{mean_j}$, with $j=1,2,3$, determines the mean of the $j$-th} state of the nodes of the system in a way that:
{
\begin{equation}\label{eq:syncro_def3}
\x_{mean_1}(t) = N^{-1}(x_{11}(t)+ x_{21}(t)+\ldots + x_{N1}(t))
  \end{equation}
}
\noindent for some $\epsilon_{max} > \epsilon > 0$.
\end{definition}
%

The upper limit $\epsilon_{max}$ will be considered for this particular arrange of the network as a fix value calculated as the average of the error $ \epsilon_{max} = N^{-1}\sum_{j=c_{min}}^{c_{max}}e_i(c_j)$ considering the variation of the coupling strength from a minimum value $c_{min}$ up to a maximum value $c_{max}$. These values will be described in the following sections. {If $\lim_{t\to \infty}\epsilon \to 0$ in \eqref{eq:syncro_def}, then the dynamical network \eqref{eq:dyNet} is said to achieve synchronous collective behavior. In experiments like electronic implementations, it is almost impossible to obtain synchronous collective behavior due to tolerance in physical devices}
%

In this paper  the practical synchronization of the dynamical network \eqref{eq:dyNet} is studied for the case in which nodes are almost identical in the sense that their difference lies in its parameters values. In the next section it is stated in detail the research problem addressed in the investigation.

\section{Problem statement}

In order to model the specific tolerances in the nominal values of our circuit's components it is introduced in the dynamical network \eqref{eq:dyNet} a parameter mismatch in each node, considering
a  variation from the original systems values simulating a maximal nominal  tolerance in electronic devices. Additional, it is assumed that the dynamics of each hybrid node is given by a PWL system based on a UDS of the Type I of the form \eqref{eq:PWL}-\eqref{eq:uds} with a switching law \eqref{eq:switching_law_2} as described in Definition \ref{Definition:1}. With these considerations, the state equation of the entire network is given by:
\begin{equation}\label{eq:Net_pmismatch}
\dot{\x}_{i}(t) = \alpha_{i}\left( \mathbf{A}\mathbf{\x}_{i} + \mathbf{B}(\mathbf{\x}_{i})   + c_{i}\sum_{j = 1}^{N} \Delta_{ij} \Gamma(\mathbf{\x}_{j}(t) -\mathbf{\x}_{i}(t)) \right), \quad i = 1,\ldots,N;
\end{equation}

\noindent where $ \alpha_{i}  \in \mathbf{R}^{+}$ with  $ i = 1,\ldots,N$ represents the parameter mismatch value for each node, considering
a  variation from the original systems value simulating a maximal nominal  tolerance in electronic devices. Notice that $\alpha_{i} $ also affects the coupling term  $c_{i}$ as $c_i\cdot \alpha_{i}$, anticipating that the physical interconnections in the network will also be implemented electronically.

\begin{figure}[!t]
\centering
\includegraphics[width=0.495\textwidth]{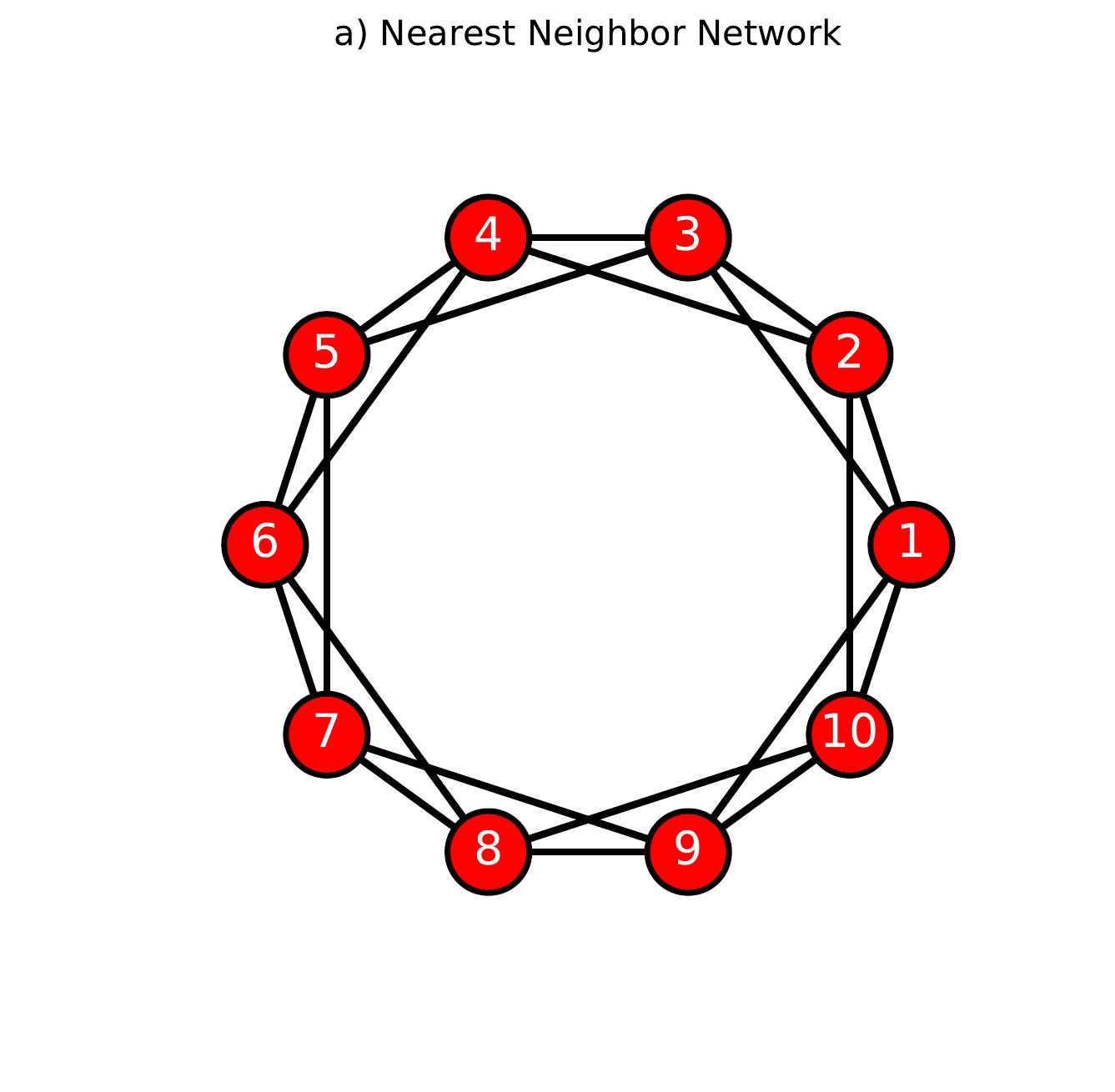}
\includegraphics[width=0.495\textwidth]{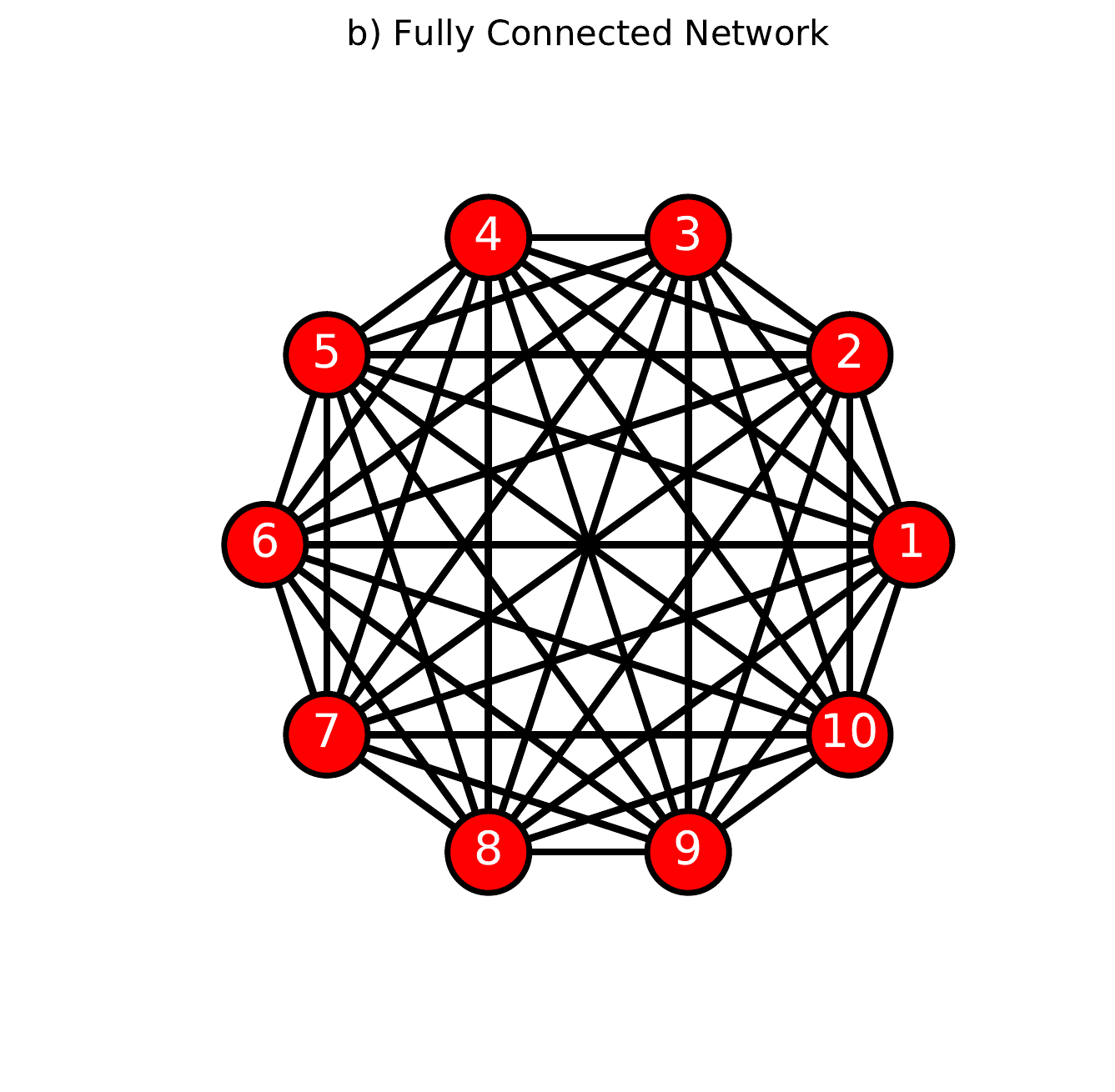}
\caption{Graph structures for the network connection of $N=10$ non-identical nodes for the numerical simulation: a) nearest neighbor (NN) connection and b) fully connected (FC) network.}
\label{fig:Grafos}
\end{figure}

Although  physical components  which are regularly manufactured present tolerances between $\pm5$\% to $\pm10$\% \cite{RobertSpence1997}. Here, higher values than the ones reported by the manufacturer will be implemented in a way that any possible deviation from the normal range of real values is considered. Then, from a random uniform distribution the  $\alpha_{i}$ variation will be given by the Matlab command $rand()$ considering values between  $\alpha_i\in [0.8,1.2)$ corresponding to  a $\pm20 \%$ random tolerance.

The research problem tackled in this paper is about  verifying whether  a set of interconnected PWL systems based on UDS given by Eq. \eqref{eq:Net_pmismatch} with parameter mismatch given by $\alpha_i$ achieve practical synchronization. The methodology consists of two steps: i) implement numerical simulations of the network model given in Eq. \eqref{eq:Net_pmismatch} and; ii) the development of an electronic circuit experiment that resembles the real physical network. In the following sections the results for each methodology are discussed.

%

\begin{figure}[t]
\centering
\includegraphics[width=0.495\textwidth]{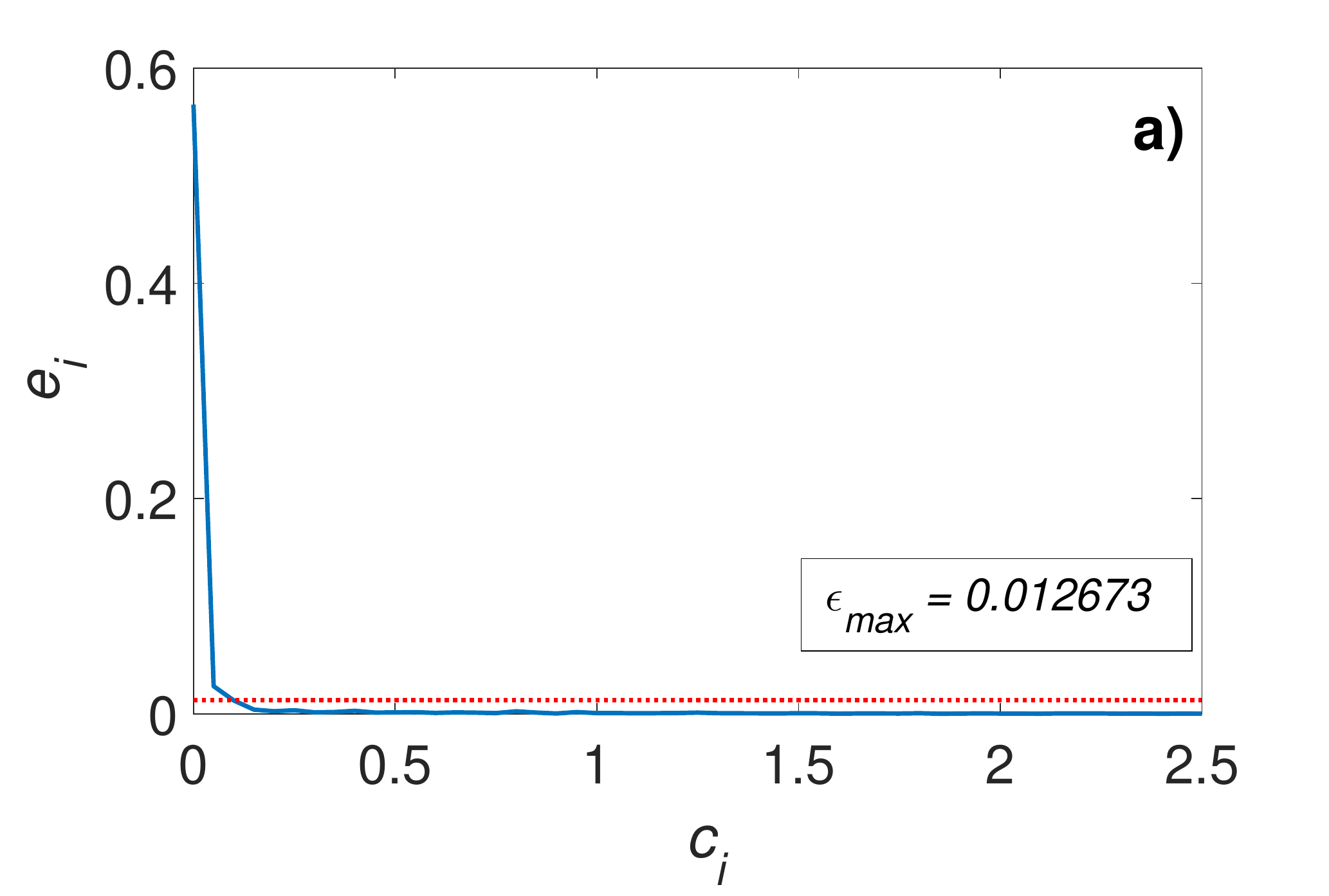}
\includegraphics[width=0.495\textwidth]{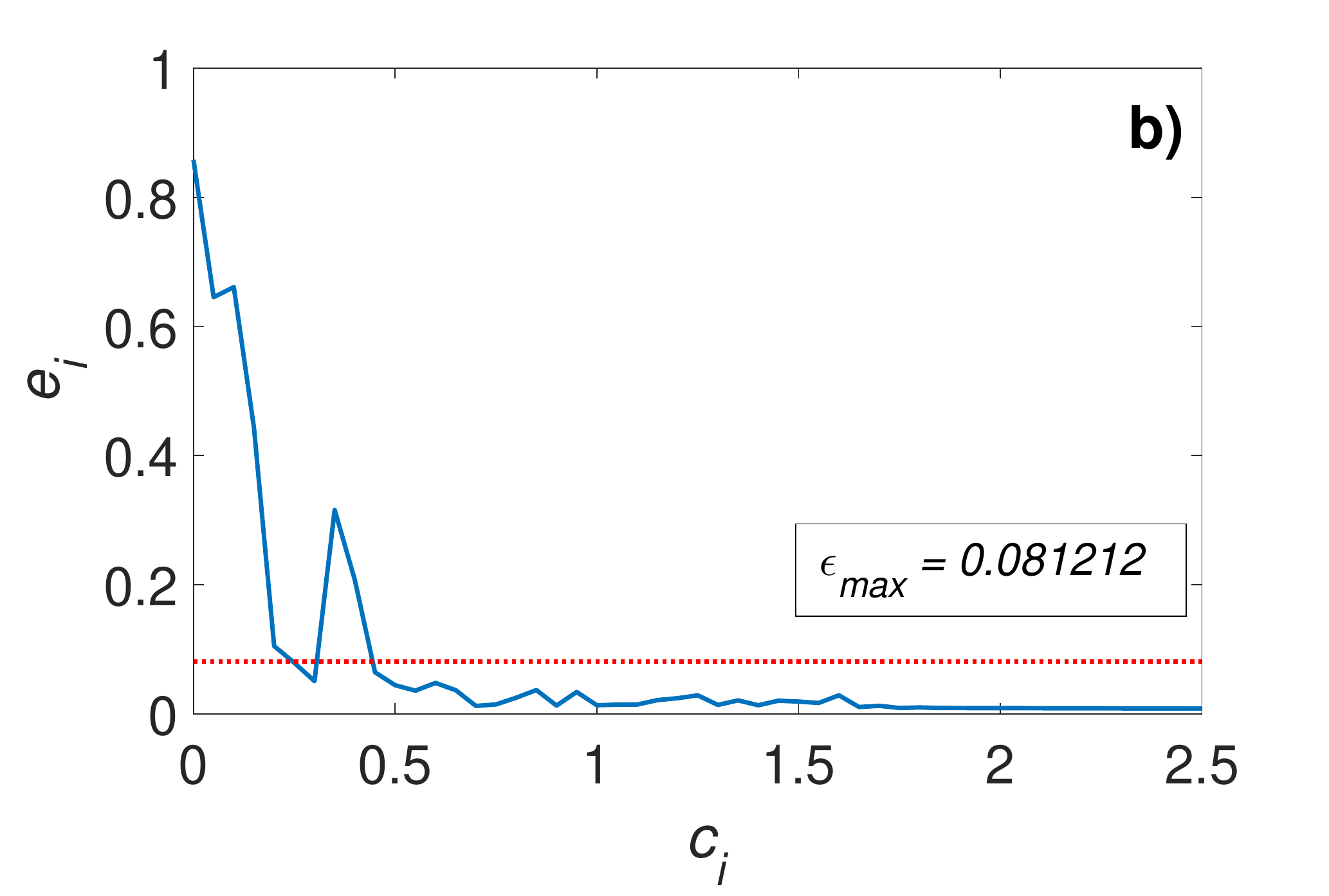}
\caption{{Euclidean} error between the states of the system nodes calculated by a variation in the coupling strength from $0<c_i<2.5$ for $N=10$ nearly identical nodes connected in: a) Fully Connected (FC) configuration b) Nearest Neighbor (NN) configuration. The red dotted line indicate the location of the $\epsilon_{max}$ value as described in Definition \ref{Definition:3}.}
\label{fig:Error_Ci}
\end{figure}

\begin{figure}[t]
\centering
\includegraphics[width=0.495\textwidth]{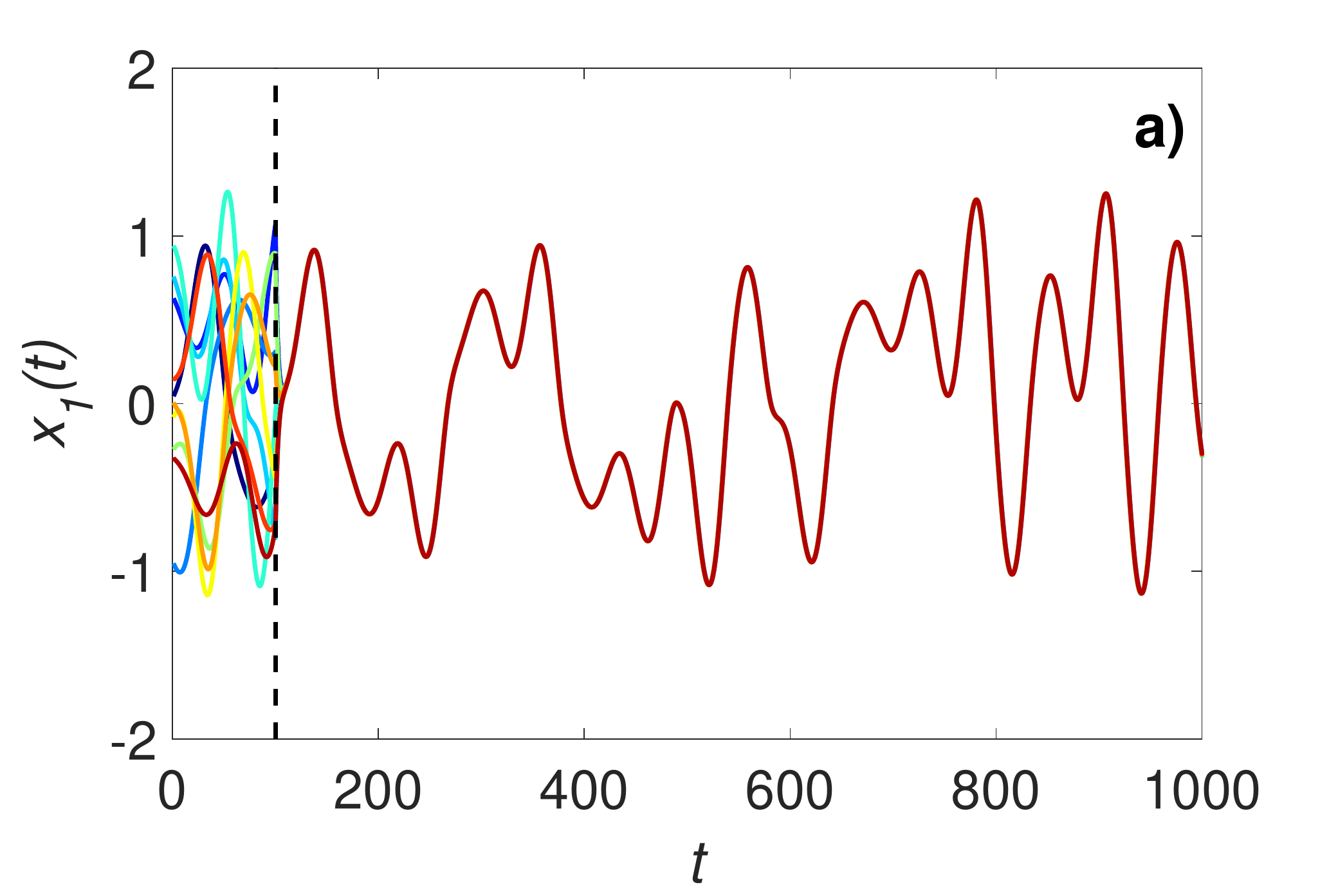}
\includegraphics[width=0.495\textwidth]{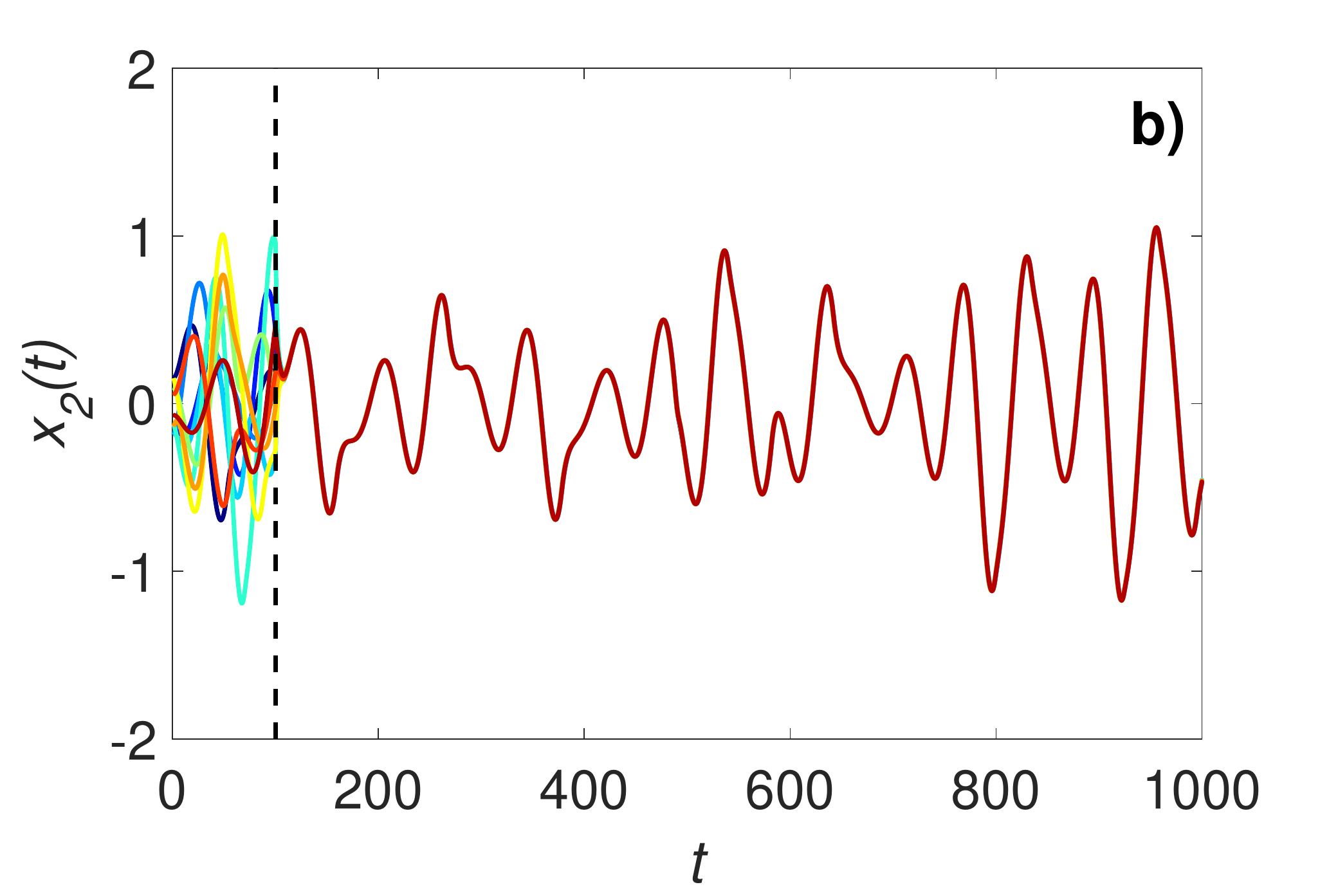}
\includegraphics[width=0.495\textwidth]{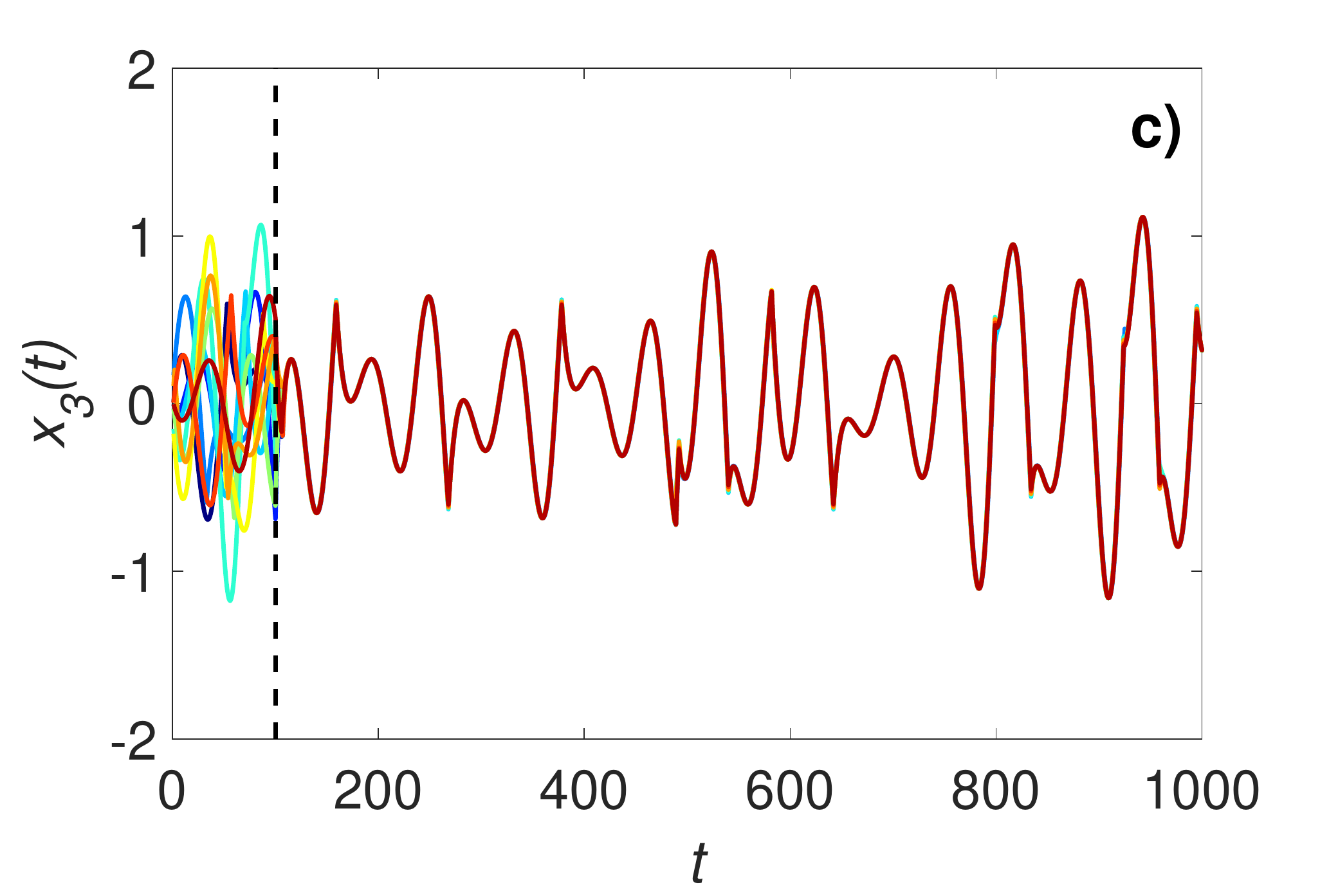}
\includegraphics[width=0.495\textwidth]{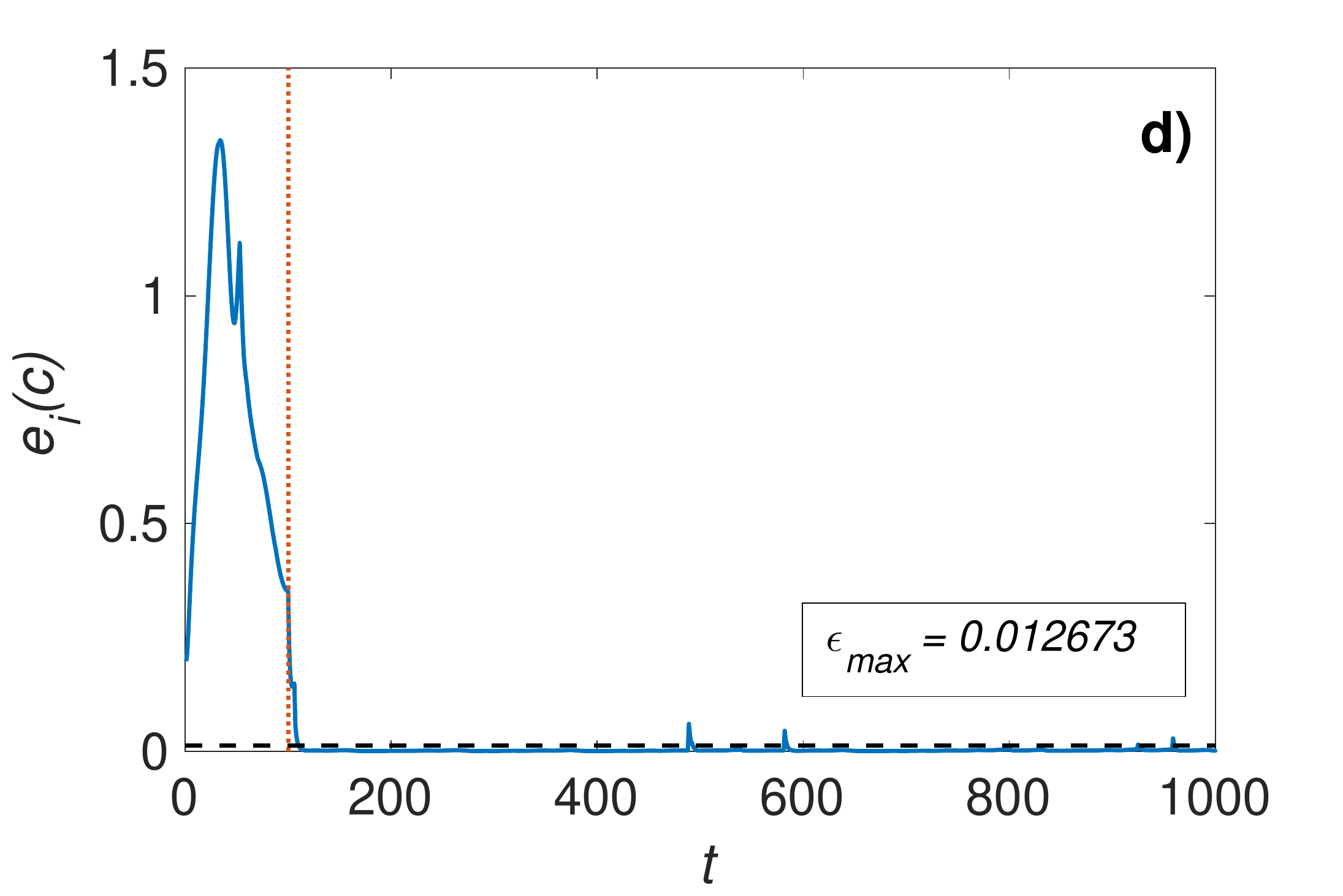}
\caption{Time Series of the variable state {a) $x_{i1}$, b) $x_{i2}$ and c) $x_{i3}$ for  $i=1,\ldots, 10$} nearly identical nodes connected in a Fully Connected (FC) configuration (Eq. \eqref{eq:FC}) with $c=0.5$. The { Euclidean} error $e_i(c)$ among the variables states of the system in node 1 is appreciated in d).}
\label{fig:FCnumerical}
\end{figure}

\section{Numerical simulation}\label{sec:NumericalSimulation}

A network of $N=10$ nearly identical nodes will be considered and connected in two network topologies:  Nearest Neighbor (NN) and Fully Connected (FC), as depicted in Figure \ref{fig:Grafos}. The diffusive coupling matrix for the NN network with two neighbors at each side is given by:

\begin{equation}\label{eq:NN}
\Delta^{near}=\left(
\begin{array}{cccccccccc}
        -4  & 1 & 1 & 0 & 0 & 0 & 0 & 0 & 1 & 1  \\
         1  & -4 & 1 & 1 & 0 & 0 & 0 & 0 & 0 & 0  \\
         1  & 1 & -4 & 1 & 1 & 0 & 0 & 0 & 0 & 0  \\
         0  & 1 & 1 & -4 & 1 & 1 & 0 & 0 & 0 & 0  \\
         0  & 0 & 1 & 1 & -4 & 1 & 1 & 0 & 0 & 0  \\
         0  & 0 & 0 & 1 & 1 & -4 & 1 & 1 & 0 & 0  \\
         0  & 0 & 0 & 0 &  1& 1 & -4 & 1 & 1 & 0  \\
         0  & 0 & 0 & 0 & 0 & 1 & 1 & -4 & 1 & 1  \\
         1  & 0 & 0 & 0 & 0 & 0 & 1 & 1 & -4 & 1  \\
         1  & 1 & 0 & 0 & 0 & 0 & 0 & 1 & 1 & -4  \\
\end{array}
\right);
\end{equation}
And the diffusive coupling matrix for the FC network is given by:
\begin{equation}\label{eq:FC}
\Delta^{fully}=\left(
\begin{array}{cccccccccc}
        -9  & 1 & 1 & 1 & 1 & 1 & 1 & 1 & 1 & 1  \\
         1  & -9 & 1 & 1 & 1 & 1 & 1 & 1 & 1 & 1  \\
         1  & 1 & -9 & 1 & 1 & 1 & 1 & 1 & 1 & 1  \\
         1  & 1 & 1 & -9 & 1 & 1 & 1 & 1 & 1 & 1  \\
         1  & 1 & 1 & 1 & -9 & 1 & 1 & 1 & 1 & 1  \\
         1  & 1 & 1 & 1 & 1 & -9 & 1 & 1 & 1 & 1  \\
         1  & 1 & 1 & 1 & 1 & 1 & -9 & 1 & 1 & 1  \\
         1  & 1 & 1 & 1 & 1 & 1 & 1 & -9 & 1 & 1  \\
         1  & 1 & 1 & 1 & 1 & 1 & 1 & 1 & -9 & 1  \\
         1  & 1 & 1 & 1 & 1 & 1 & 1 & 1 & 1 & -9  \\
\end{array}
\right).
\end{equation}

\begin{figure}[t]
\centering
\includegraphics[width=0.495\textwidth]{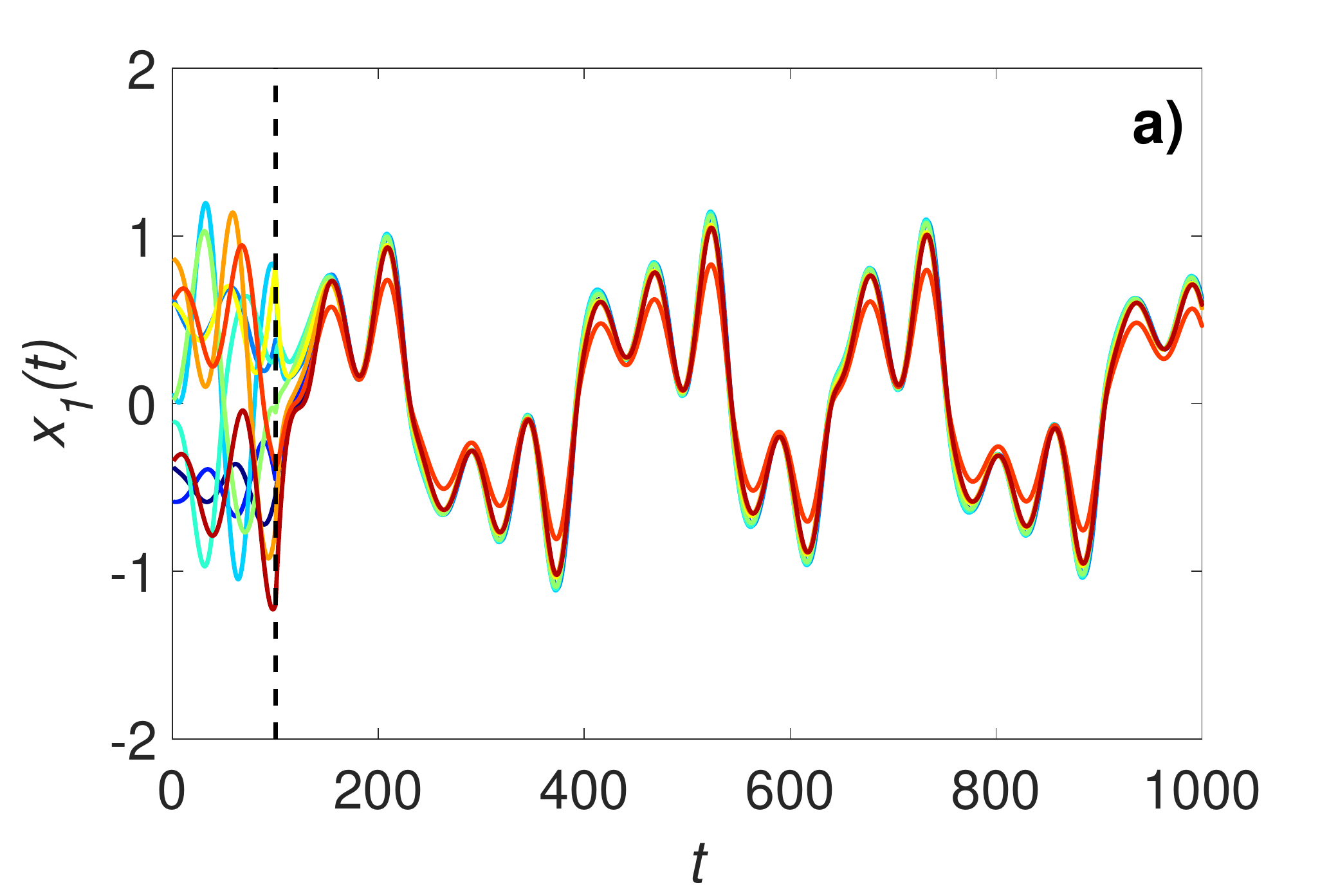}
\includegraphics[width=0.495\textwidth]{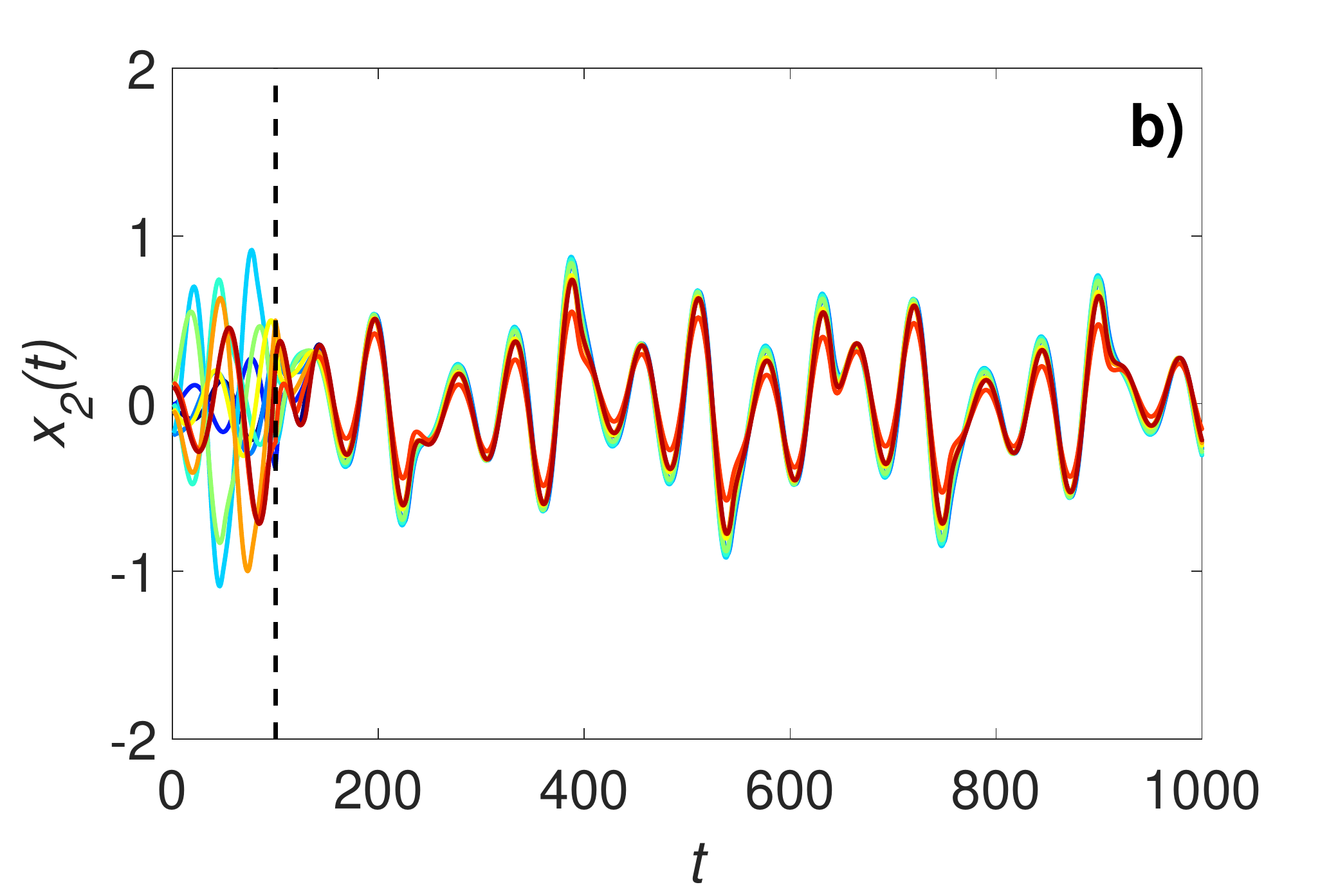}
\includegraphics[width=0.495\textwidth]{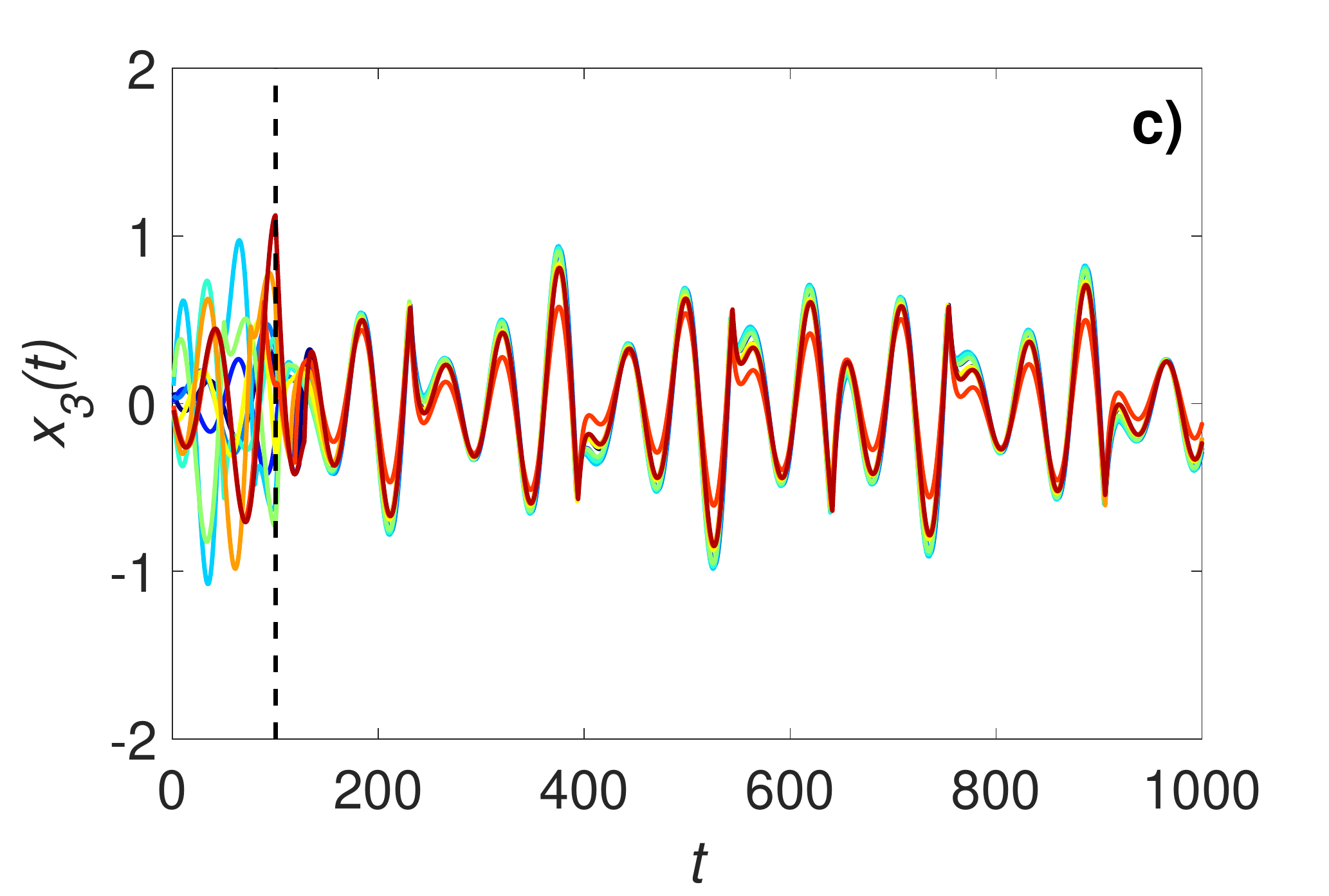}
\includegraphics[width=0.495\textwidth]{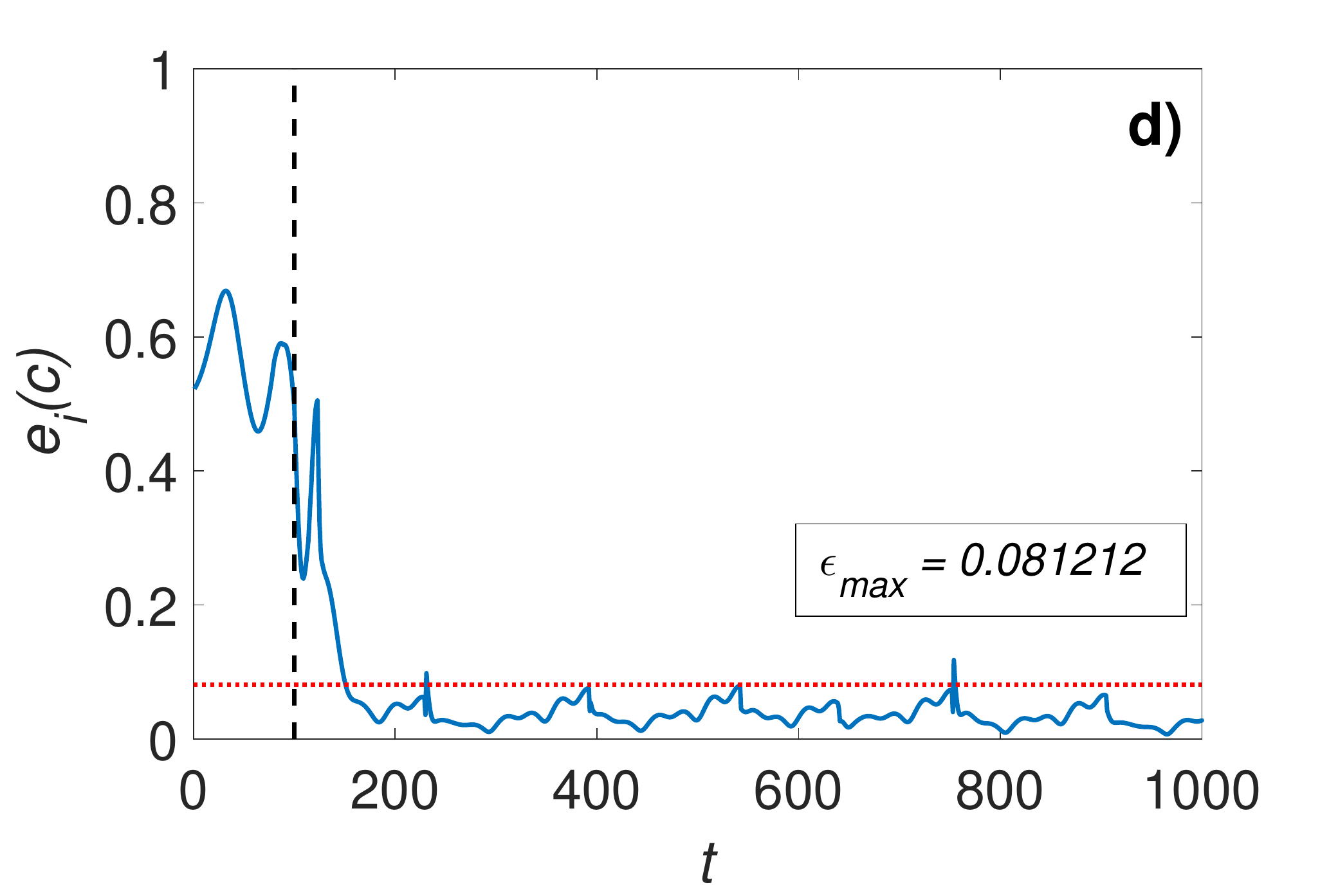}
\caption{Time Series of the variable state {a) $x_{i1}$, b) $x_{i2}$ and c) $x_{i3}$ for  $i=1,\ldots,10$} nearly identical nodes connected in a Nearest Neighbors (NN) configuration (Eq. \eqref{eq:FC}) with $c=0.5$. The {Euclidean} error $e_i(c)$ among the variables states of the system in node 1 is appreciated in d).}
\label{fig:NNnumerical}
\end{figure}

%

Regarding the link attributes, it is assumed that  the inner coupling matrix is given by the identity matrix \textit{i.e.} $\Gamma =  \mathbb{I} \in \R^{3}$. The random tolerance parameter $\alpha_i$ will take the values given in Table \ref{Table:Init_Config} where also the initial condition is shown for each one of the nodes. To determine the correct value of the coupling strength $c$ between the nodes the {Euclidean} distance as represented in Definition \ref{Definition:3} was calculated for a variation of the values in the range $0\leq c \leq 2.5$ for two different connections. The results are depicted in Figure \ref{fig:Error_Ci} a) for the FC network and in Figure \ref{fig:Error_Ci} b) for the NN network. Both graphs also depict a red dotted line indicating the  location of the $\epsilon_{max}$ value as described in Definition \ref{Definition:3}, the text boxes show that value for each type of connection. So any network node which its oscillating states present an {Euclidean} distance below the $\epsilon_{max}$ value, will be considered practical synchronized.


\begin{table}[!b]
\centering
\caption{The initial condition and the corresponding parameter mismatch for each one of the $N=10$ and connected according to a FC and NN configuration.}
\begin{tabular}{ccc}
\hline
   Node index & Initial Condition        &   $\alpha_{i}$ \\
\hline
      1 & (-0.153, 0.407, -0.388)  &         0.9369 \\
      2 & (0.107, -0.167, -0.48)   &         0.9548 \\
      3 & (0.29, 0.297, 0.049)     &         0.8104 \\
      4 & (-0.025, 0.464, 0.367)   &         1.0534 \\
      5 & (-0.121, -0.203, -0.026) &         1.0261 \\
      6 & (0.264, -0.285, -0.3)    &         0.9462 \\
      7 & (0.065, -0.407, -0.208)  &         0.9685 \\
      8 & (0.084, -0.223, 0.355)   &         0.9848 \\
      9 & (-0.09, 0.324, -0.159)   &         0.9257 \\
     10 & (0.039, -0.012, 0.123)   &         0.8902 \\
\hline
\label{Table:Init_Config}
\end{tabular}
\end{table}

Taking this in consideration, in Figures \ref{fig:FCnumerical} a),b) and c) it is shown the dynamics of the nearly-identical network \eqref{eq:Net_pmismatch} with a FC configuration \eqref{eq:FC}. It can be appreciated how {the $x_{i1},x_{i2}$ and $x_{i3}$ states of the $10$ nodes $(i=1,\ldots,10)$} are oscillating autonomously until the network is coupled in the nearest neighbor connection at $t \geq 100$, which represents the moment in which the coupling strength changes from $c = 0$ to $c = 0.5$.  The solutions of the systems were calculated  with the Runge-Kutta of the fourth order integration method and with an integration step of 0.01. Notice how the states begin to couple among themselves presenting some transient behavior after $t \geq100 $. Also, the {Euclidean} error is shown for the variable state of the node $1$ in \ref{fig:FCnumerical} d). Here it can be seen how error fall for $t>100$ below a value of $\epsilon_{max}$, that is, the network  \eqref{eq:Net_pmismatch} {achieves  practical synchronous collective behavior.}

Now, the case of the NN network configuration  whose diffusive coupling matrix  is given by \eqref{eq:NN} is shown in Figures \ref{fig:NNnumerical} a), b), c) and d). Here, the time series of each variable and the corresponding euclidian error is depicted, similar characteristics are considered as in the previous example, i.e.,  at $t \geq 100$ the coupling strength changes from $c = 0$ to $c = 0.5$. It is shown that the nearly-identical network \eqref{eq:Net_pmismatch}  {achieves practical synchronous collective  behavior} considering $\epsilon_{0.81212}$, except for some brief instants in which the system present intermittency on the synchronous states after $t=250$ time steps. This intermittency phenomena happens because of the less connections among the nodes that the network presents, in contrast as in the FF the coupling. To avoid redundancy in the results depicted the only value of the coupling strength that is presented is at $c = 0.5$, since for greater values as depicted in Figure \ref{fig:Error_Ci} the error falls below $\epsilon_{max}$.

In the next section the electronic circuit implementation of the nearly-identical network \eqref{eq:Net_pmismatch} will be described to study experimentally the bounded synchronization phenomenon.

\section{Electronic circuit implementation}

In order to demonstrate the network coupling in a real physical manner, the electronic implementation of the network {is} carried out. Following the concepts of analog computation described in \cite{Orponen1997} and in the same {way that} \cite{AnalogUDSI}, these systems can be electronically implemented by means of specific configurations of Op Amps. The idea is described next.

A network  interconnected according to the coupling structures in Figure~\ref{fig:Grafos} will be considered, with the variation that it will be composed of $N=4$ nodes.  The systems states will be connected among each others according to the state linking matrix $\Gamma= \mathbf{I}_{3}$ where $\mathbf{I}\in R^3$ represents the identity matrix. The coupling strength between connections will be implemented by $c_i$.
Now, in order to represent the electronic implementation of one of the nodes, firstly it will be considered the {$x_{11}$ and $x_{12}$} state equations (for the first node, {\it i.e.} $N=1$), represented in \eqref{eq:PWL}-\eqref{eq:uds} with \eqref{eq:switching_law_2} coupled to the network by means of \eqref{eq:Net_pmismatch}. The set of equations for the $i-th$  node  will take the following form:

\begin{equation}\label{ec_node1}
\mathbf{\dot{x}}_i= \left[
    \begin{array}{l}
      \dot{x_{i1}}\\
      \dot{x_{i2}}\\
      \dot{x_{i3}}
   \end{array}
 \right] = \left[
    \begin{array}{l}
        x_{i2} + c\sum_{j=1}^{3}\Delta_{ij}(x_{j1}-x_{i1})\\
         x_{i3} + c\sum_{j=1}^{3}\Delta_{ij}(x_{j2}-x_{i2})\\
        -2x_{i1}-x_{i2}- x_{i3}+b_{i3}(\x) +$ $c_i\sum_{j=1}^{3}\Delta_{ij}(x_{j3}-x_{i3})
   \end{array}
\right].
\end{equation}

\begin{figure}[!t]
\centering
\includegraphics[width=11cm]{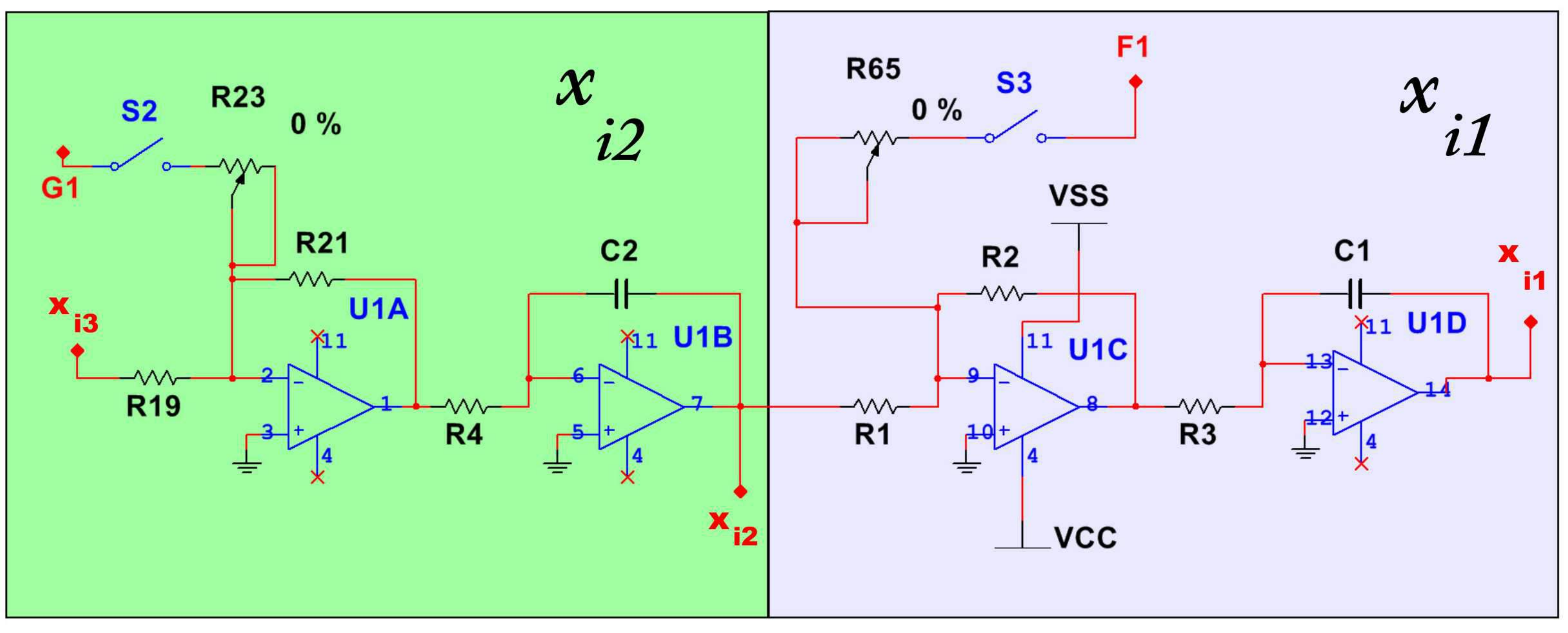}
\caption{\label{fig:node1_x1y1} Circuit diagram for the electronic implementation of the $x_{i1}$ and $x_{i2}$ state equations of the $i-th$ node,  as described in eq. \eqref{ec_node1_int}.}
\end{figure}

Notice that the parameter $\alpha_i$ has not been considered explicitly in the equation \eqref{ec_node1} because it is considered implicitly in the electronic circuit with the corresponding tolerance naturally in their physical devices. Now, by  integrating with respect to time on both sides of the equations, the states will result in:

\begin{equation}\label{ec_node1_int}
\left[
    \begin{array}{l}
      {x_{i1}(t)}\\
      {x_{i2}(t)}\\
      {x_{i3}(t)}
   \end{array}
 \right] = \left[
    \begin{array}{l}
        \int_0^t{\big(  x_{i2}(\tau) + c_i\sum_{j=1}^{3}\Delta_{ij}(x_{j1}(\tau)-x_{i1}(\tau)) \big)}d\tau\\
         \int_0^t{\big(x_{i3}(\tau) + c_i\sum_{j=1}^{3}\Delta_{ij}(x_{j2}(\tau)-x_{i2}(\tau)) \big)}d\tau\\
			\int_0^t \big(-2x_{i1}(\tau)-x_{i2}(\tau)- x_{i3}(\tau) + b_{i3}(\x(\tau))+\\
				 c_i\sum_{j=1}^{3}\Delta_{ij}(x_{j3}(\tau)-x_{i3}(\tau)) \big) d\tau
   \end{array}
\right].
\end{equation}

\begin{figure}[!t]
\centering
\includegraphics[width=11cm]{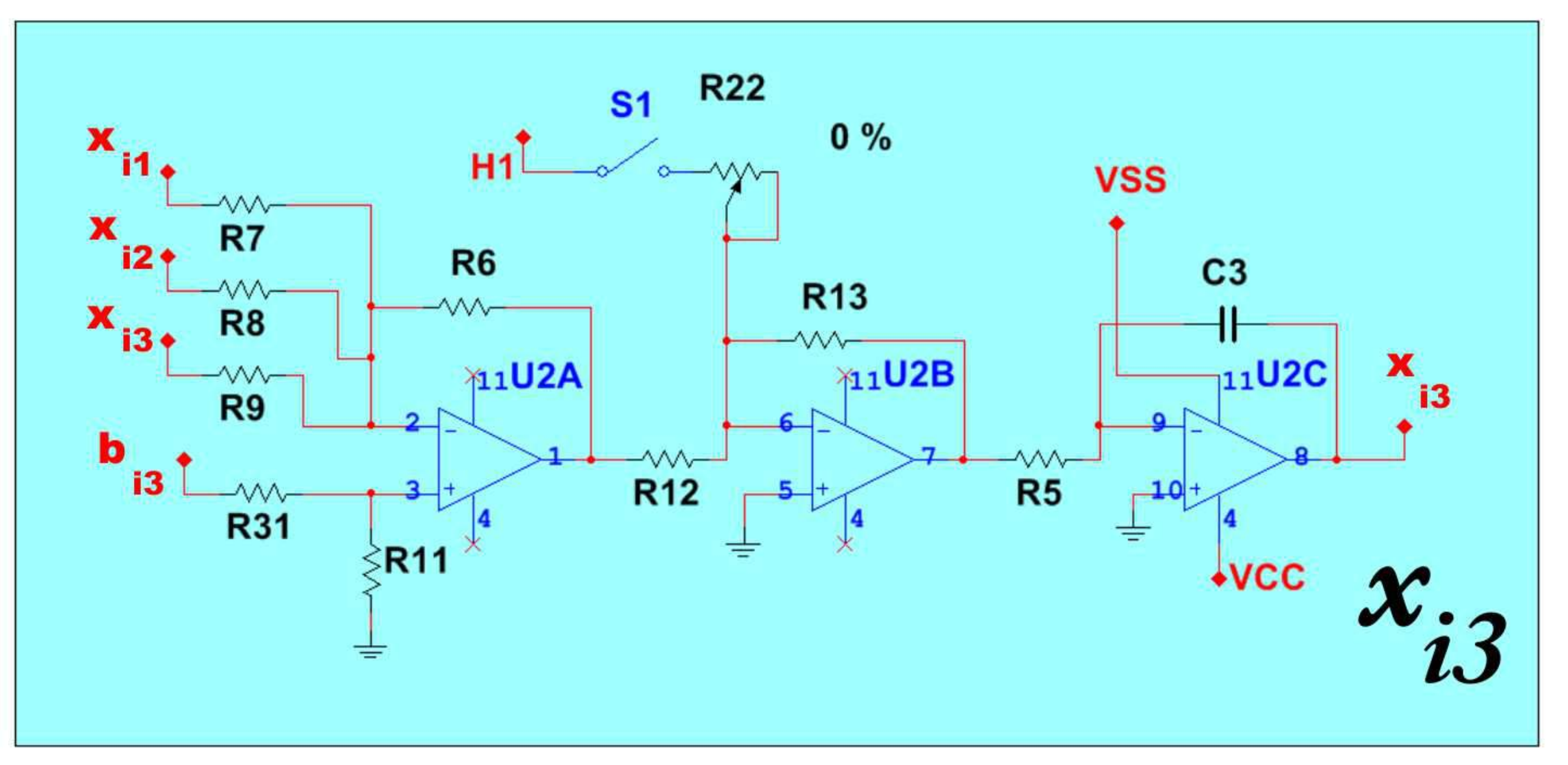}
\caption{\label{fig:node1_z1} Circuit diagram for the electronic implementation of the $x_{i3}$ state equation of the node $N=1$,  as described in eq. \eqref{ec_node1_int}.}
\end{figure}
\begin{figure}[!t]
\centering
\includegraphics[width=8cm]{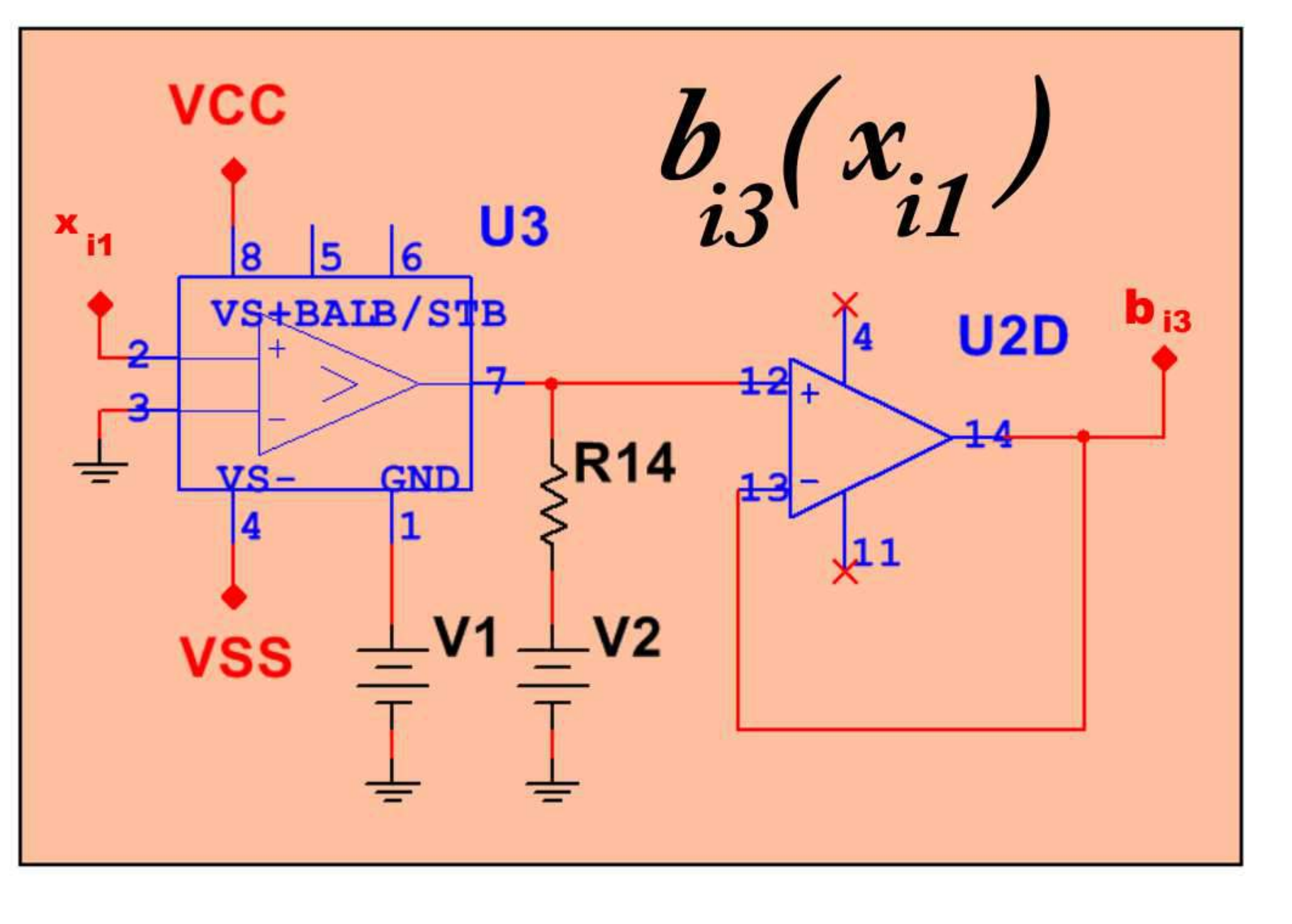}
\caption{\label{fig:node1_B} Circuit diagram for the electronic implementation of the commutation values of the node $N=1$,  as described in eq. \eqref{ec_node1_int}.}
\end{figure}

By means of the integration configuration in the Op Amps with the inverting capacitor feedback connection, one can model electronically the state equations in matter. This can be appreciated for the electronic implementation of one single node given by the eq. \eqref{ec_node1} with \eqref{ec_node1_int}, depicted in Figure \ref{fig:node1_x1y1} for each of the corresponding states $x_{i1}$ and $x_{i2}$, in Figure \ref{fig:node1_z1} for $x_{i3}$, and in Figure \ref{fig:node1_B} for the switching value of $b_{i3}$ depending on the value of $x_{i1}$. By using common node analysis by Kirchhoff current law and the superposition technique in the nodes marked as $x_{i1}, x_{i2}, x_{i3}$ in the circuit implementation mentioned before, the following voltage equation will result.

\begin{equation}\label{ec_voltage}
    \begin{array}{lll}
      {x_{i1}}& =&  \frac{-1}{R3\cdot C1}\int \bigg(  -\frac{R2}{R1}x_{i2}  - \frac{R2}{R65}S1\cdot F1 \bigg)dt-V_{C1_{0}},\\
      {x_{i2}}& =&  \frac{-1}{R4\cdot C2}\int \bigg(  -\frac{R21}{R19}x_{i3} - \frac{R21}{R23}S2\cdot G1 \bigg)dt-V_{C2_{0}},\\
      {x_{i3}}& =&  \frac{-1}{R5\cdot C3}\int \bigg(\frac{R6\cdot R13}{R7\cdot R12}x_{i1} + \frac{R6\cdot R13}{R8\cdot R12}x_{i2} + \frac{R6\cdot R13}{R9\cdot R12}x_{i3}\\
      &&- \frac{R13}{R22}H1-\frac{R11\cdot R13\cdot(RP+R6)}{RP\cdot R12\cdot(R31+R11)}b_{i3} \bigg) dt-V_{C3_{0}};
    \end{array}
\end{equation}

\noindent where $RP=R7\|R8\|R9$, with ``$\|$'' as the equivalent parallel resistor.
In the three state circuits there is a connection represented by  the switch components S1, S2, and S3, with which one is able to implement the ``$0$'' or ``$1$'' values from the inner linking matrix $\Gamma$  connections. The coupling strength of the network $c$ is represented by means of the relation throughout the resistor and potentiometers marked as R2, R13, R21, R22, R63, R65  in a way that $c =\frac{R2}{R65}=\frac{R21}{R23}=\frac{R13}{R22}$. Therefore, the value of the coupling strength can be varied by the adjustment of the variable resistors depending of the experiment requirements. The commutation values that the system has are represented by the comparator amplifier in Figure \ref{fig:node1_B}. Where the changing of sign of the signal $x1$ is being implemented by the U3 component according to the commutation surface in the systems equation given in \eqref{eq:switching_law_2}.

\begin{figure}[!t]
\centering
\includegraphics[width=13cm]{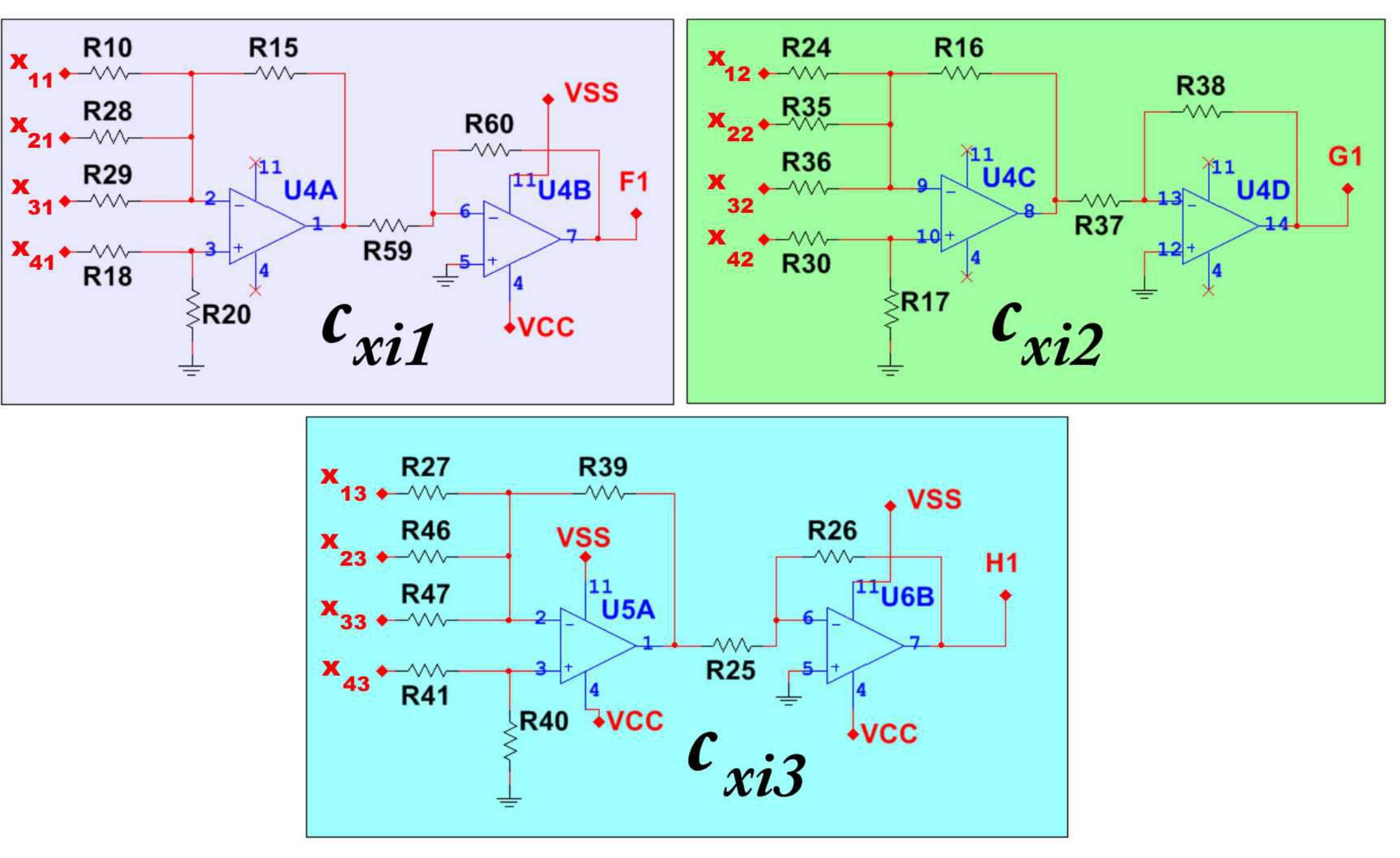}
\caption{\label{fig:node1_coupling} Circuit diagram for the electronic implementation of the coupling of the node $N=1$,  as described in eq. \eqref{ec_node1_int} by the sum term with eq. \eqref{ec_voltage2} and a matrix connection $\Delta^{fully}$  as depicted in eq. \eqref{eq:FCcircuito}.}
\end{figure}

The term of the sum involved in the coupling in each of the state equations in \eqref{ec_node1} with \eqref{ec_node1_int} is performed by the circuit displayed in Figure \ref{fig:node1_coupling}. Notice that for each of the circuits here $c_{x_{i1}},c_{x_{i2}}$ and $c_{x_{i3}}$, there are four signals connected in the inverting adding amplifiers  U4A, U4C and U5A, marked as $x_{ij}$ with $i,j=1,2,3,4$, respectively. The nodes marked as $F1, G1$ and $H1$ present the following voltage equations:

\begin{equation}\label{ec_voltage2}
    \begin{array}{lll}
      {F1}& =&  \frac{R60}{R59} \bigg(  \frac{R15}{R10}x_{41}  + \frac{R15}{R28}x_{31}+ \frac{R15}{R29}x_{21}
                -\frac{R20\cdot(RPx+R15)}{RPx\cdot (R20+R28)}x_{11} \bigg),\\
      {G1}& =&  \frac{R38}{R37} \bigg(  \frac{R16}{R24}x_{42}  + \frac{R16}{R35}x_{32}+ \frac{R16}{R36}x_{22}
                -\frac{R17\cdot(RPy+R16)}{RPy\cdot (R18+R20)}x_{12} \bigg),\\
      {H1}& =&  \frac{R26}{R25} \bigg(  \frac{R39}{R27}x_{43}  + \frac{R39}{R46}x_{33}+ \frac{R39}{R47}x_{23}
                -\frac{R40\cdot(RPz+R39)}{RPz\cdot (R40+R41)}x_{13} \bigg);\\
    \end{array}
\end{equation}

\noindent where $RPx=R10\|R28\|R29$, $RPy=R24\|R35\|R36$ and $RPz=R27\|R46\|R47$.   This node voltage equation will result in the same equation as the one given by the sum coupling term in \eqref{ec_node1}  and $\Gamma = diag\{\mathbf{I}_3 \}$. With this connections for each of the corresponding states,  the circuit will be ready to couple in the network in different configurations. It is important to mention that in case that other type of coupling is considered, the connections on the resistors ($R10, R18, R28, R29$ for $x$, $R24, R30, R35, R36$ for $x_{i2}$ and $R27, R41, R46, R47$ for $x_{i3}$), must be disconnected and the remaining resistors adjusted in value in order to represent both the connection of the matrix coupling $\mathbf{S}$ and the inner linking matrix $\Gamma$.

The values of the electronic components (resistors, capacitors and op amps)  given in the electronic implementation are depicted at the Table \ref{tab:valores2}. This values were considered so that the voltages depicted in eq. \eqref{ec_voltage} become analytically equal to the equation of the first node given by \eqref{ec_node1}. It is important to remark that the diagram circuits depicted here are considering only the first node of the network. For each other node connected in the network, a corresponding diagram from Figures \ref{fig:node1_x1y1} to \ref{fig:node1_coupling}) must be design according to the parameters of the system, and the connection and variables involved in the coupling, resulting in a large number of components and connections for a 4-node network.

\subsection{Electronic implementation of a nearest neighbor connected network}

\begin{figure}[!t]
\centering
\includegraphics[width=0.45\textwidth]{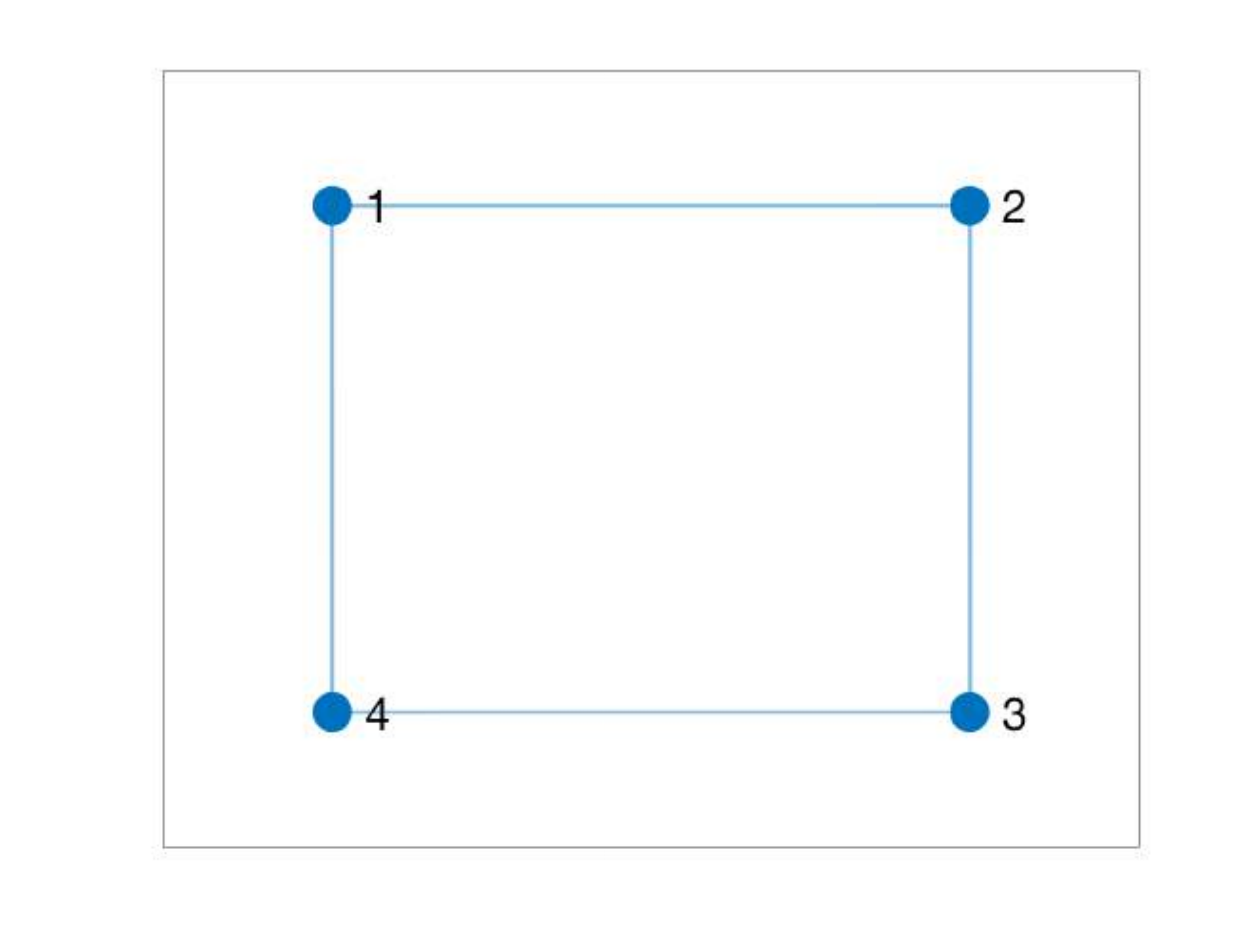}
\includegraphics[width=0.45\textwidth]{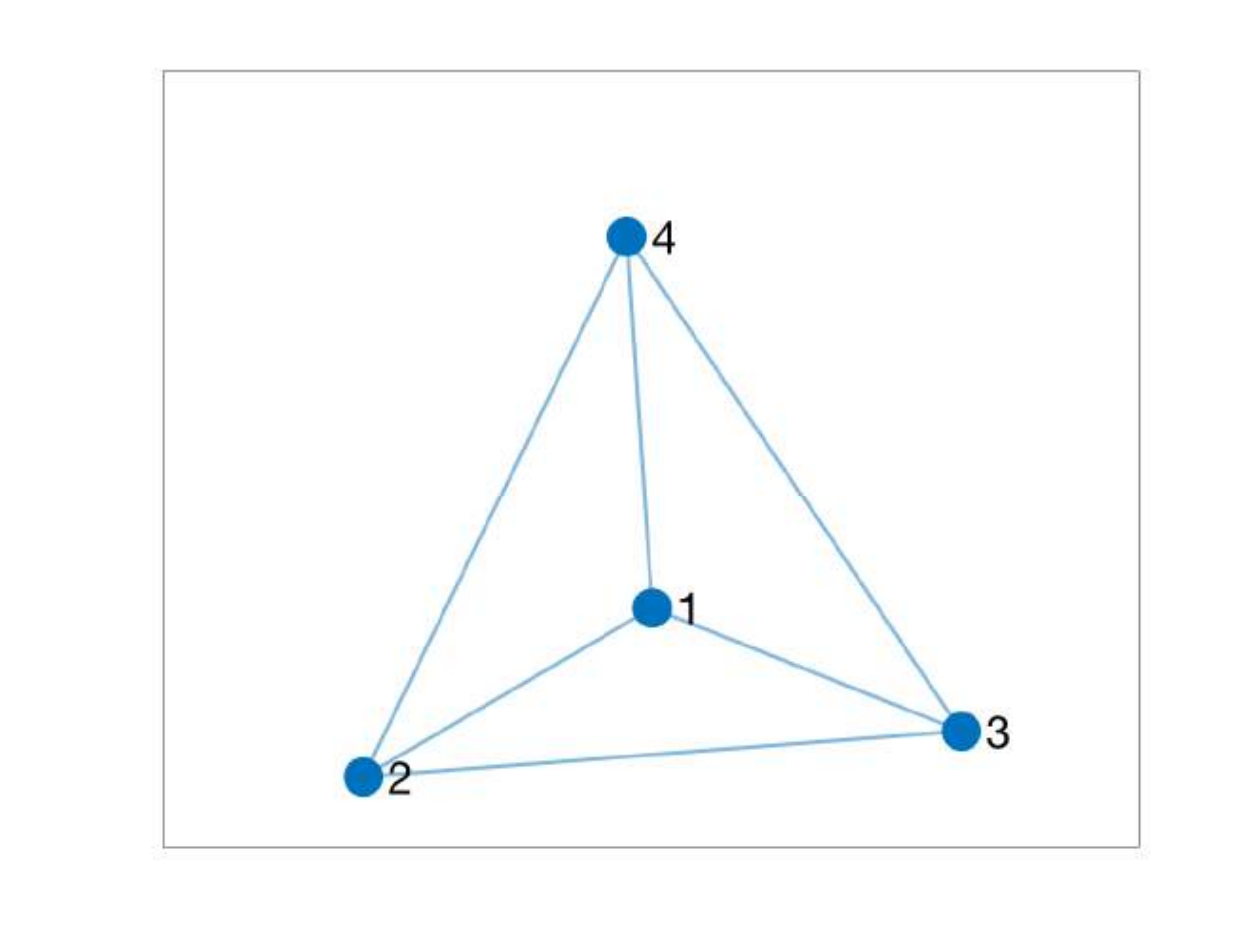}
\caption{\label{fig:Grafos_circuito} Graph structures for the electronic implementation of the network connection of $N=4$ non-identical nodes. In the left a nearest neighbor connection and in the right  a fully connected network.}
\end{figure}

In order to represent the two types of network connection given in Section \ref{sec:NumericalSimulation}, four nodes connected according to the two graphs depicted in Figure \ref{fig:Grafos_circuito} will be considered. The coupling matrices of both networks satisfying the diffusive condition will be given  by:

\begin{equation}\label{eq:FCcircuito}
\Delta^{near}=\left(
\begin{array}{cccccccccc}
        -2  & 1  & 0 & 1  \\
         1  & -2 & 1 & 0  \\
         0  & 1 & -2 & 1  \\
         1  & 0 & 1 & -2
\end{array}
\right);
\Delta^{fully}=\left(
\begin{array}{cccc}
        -3  &  1 & 1 & 1    \\
         1  & -3 & 1 & 1    \\
         1  &  1 & -3 & 1   \\
         1  &  1 & 1 & -3
\end{array}
\right);
\end{equation}

\noindent with $\Delta^{near}$ for a NN network and $\Delta^{fully}$ for a FC network. First consider the case for the nearest neighbor connection. Here the coupling must be adjusted in the node 1 to connect only with node 2 and node 4. Therefore resistors $R28, R35, R46$ which connect to the states of node 3 will be disconnected. In the same way, node 2 must present resistors connecting to node 1 and node 3, node 3 connected to node 2 and node 4, and node 4 connected to node 1 and node 3. The values of the resistors for Figure \ref{fig:node1_coupling} considering this type of coupling are given in Tables \ref{tab:valores2} and \ref{tab:valores3}.

\begin{table}[!b]
  \centering
  \caption{\label{tab:valores2} Values and components of the electronic implementation.
Any resistance not included here (considering also the potentiometers) or in the Table \ref{tab:valores3} is consider with a value of 10k$\Omega$.}
\begin{tabular}{||c|c||c|c|}
\hline
Component & Value or name & Component & Value or name\\
\hline
R3,R4,R5 & 1k$\Omega$ &  U1-U2 \& U4-U6 & TL084CD\\
R7& 5k$\Omega$ & U3 & LM319N\\
R31& 40k$\Omega$ &C1-C3 & 1$\mu$F\\
VCC & 18V &VSS & -18V\\
V1  & -1V &V2  & 1V\\
\hline
\end{tabular}
\end{table}
%

\begin{figure}
    \centering
    \begin{subfigure}[t]{0.47\textwidth} 
    \caption{}
        \includegraphics[width=\textwidth]{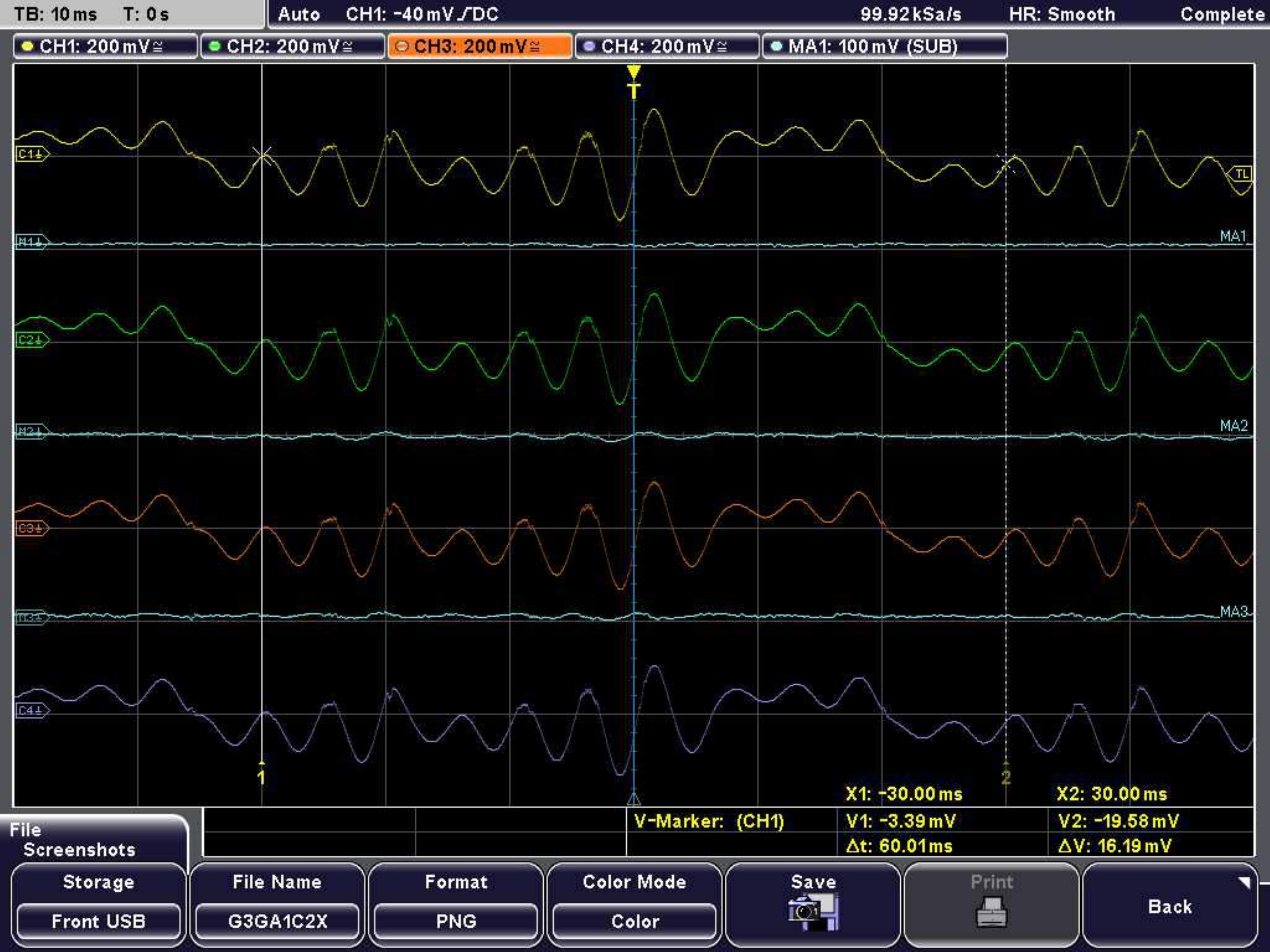}
    \end{subfigure}
    \begin{subfigure}[t]{0.47\textwidth} 
    \caption{}
         \includegraphics[width=\textwidth]{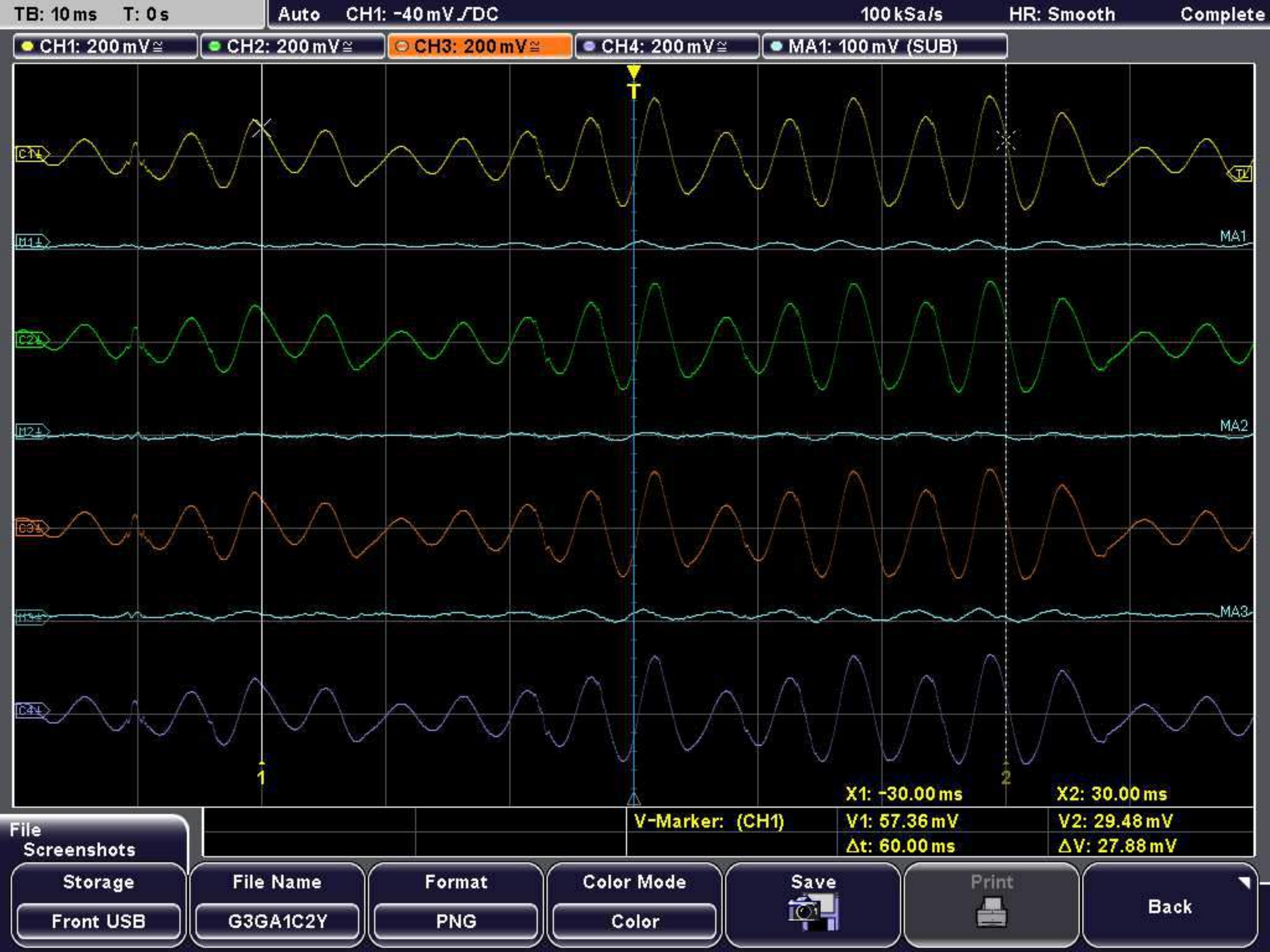}
    \end{subfigure}
        \begin{subfigure}[t]{0.47\textwidth} 
        \caption{}
        \includegraphics[width=\textwidth]{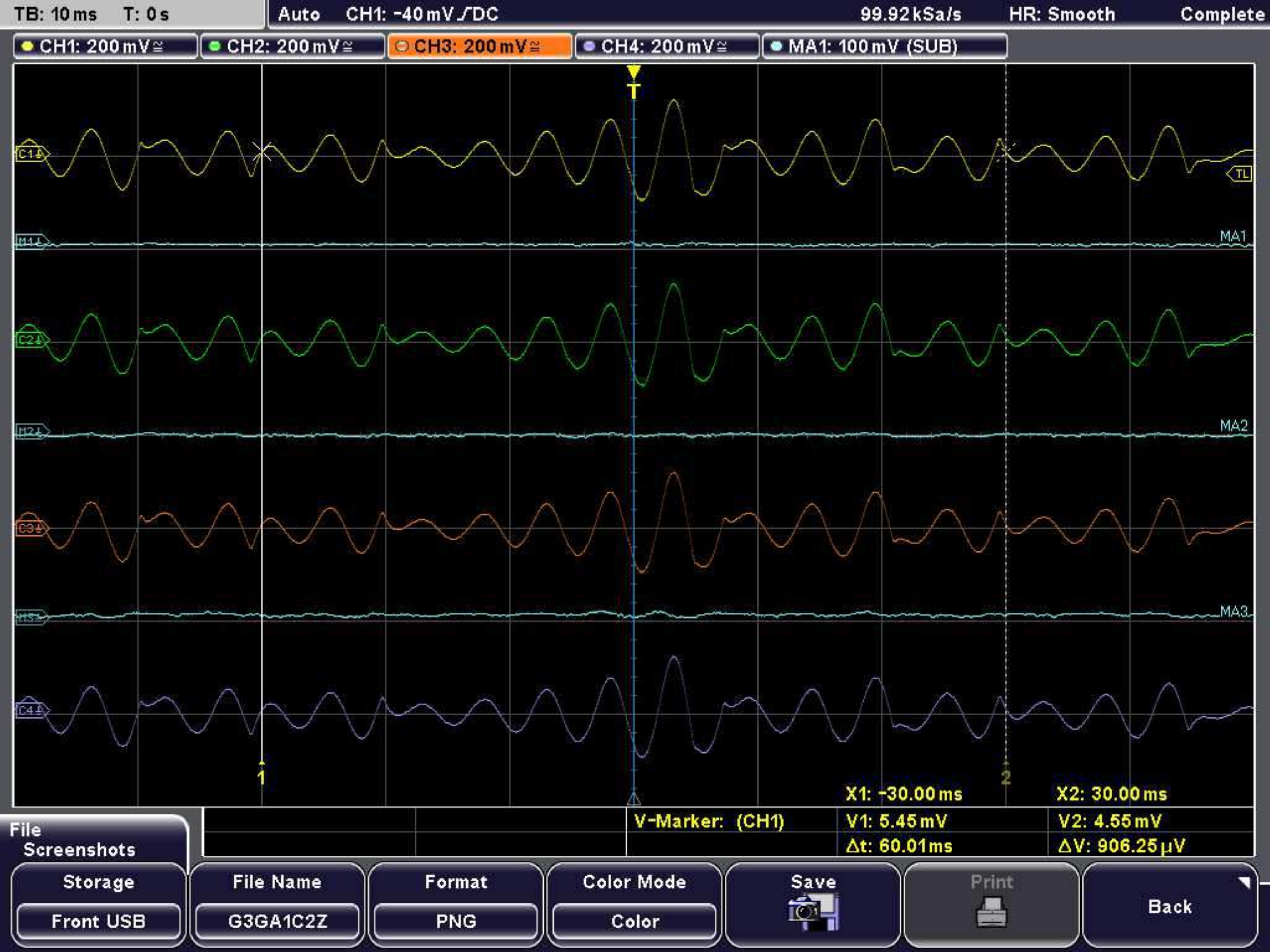}
    \end{subfigure}
    \begin{subfigure}[t]{0.47\textwidth} 
    \caption{}
         \includegraphics[width=\textwidth]{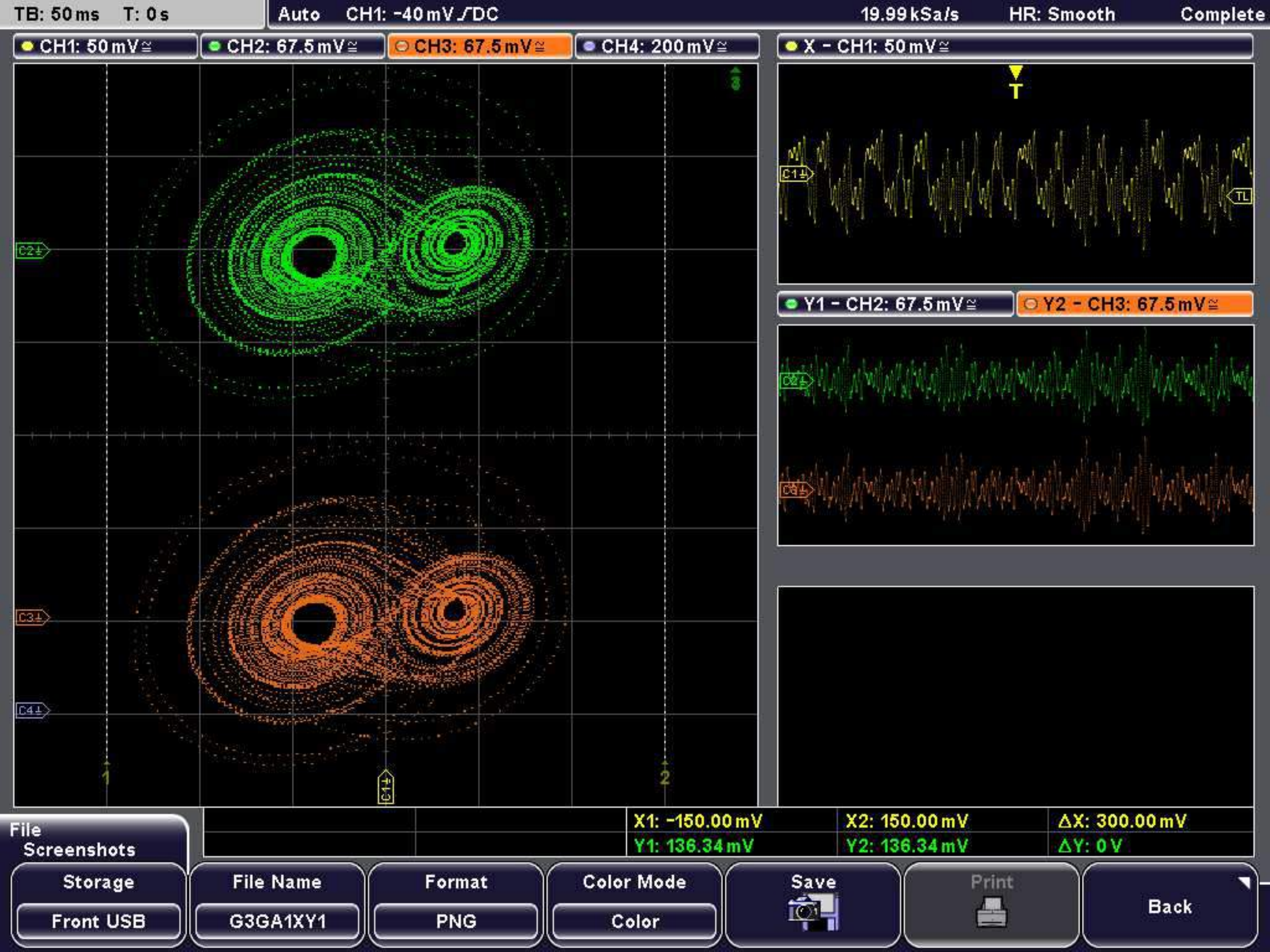}
    \end{subfigure}
\caption{Experimental traces from the oscilloscope for the signals of a) $x_{i1}$, b) $x_{i2}$, c) $x_{i3}$ of a $N=4$ nearest neighbor network implemented electronically in the circuits depicted from Figures \ref{fig:node1_x1y1}-\ref{fig:node1_coupling}, with $\Delta^{near}$ in eq. \eqref{eq:FCcircuito}, $\Gamma=diag{\mathbf{I}_3}$ with $c=2.5$. In d) is depicted the projection of the signals into the $(x_{11},x_{22})$ plane and the $(x_{11},x_{32})$ plane. The values for the resistor in the coupling of Figure \ref{fig:node1_coupling} are given in Tables \ref{tab:valores2} and \ref{tab:valores3}.}
\label{fig:circuit_G1x2}
\end{figure}

The experimentation of the network is measured with a Rohde \& Shwarz RTM2054 four channel Digital Oscilloscope. The results for the nearest neighbor coupling of $N=4$ with a coupling strength of $c=2.5$ are depicted in the experimental traces in  Figure \ref{fig:circuit_G1x2} a) to d). In which the experimental traces in time of the $x_{i1}$ state are given in Figure \ref{fig:circuit_G1x2} a) where each channel(CH1-CH4 marked in yellow, green, red and purple) correspond to each node (1-4) to the signals of $x_{11},x_{21},x_{31},x_{41}$, respectively. The signals marked as MA1-MA3 (located between each oscillating pair or measurement in light blue color) correspond to the mathematical operation $MA1 = CH2-CH1, MA2 = CH3-CH1, MA3 = CH4-CH1$. By means of this MA signals, one is able to measure the relation between the states to determine if they are synchronized. Similar specifications regarding Figure \ref{fig:circuit_G1x2} b) and c) for the states $x_{i2}$ and $x_{i3}$ respectively. Notice how MA1 is almost fully attenuated at zero, and notice also the voltage scale marked at $100 mV$. Small perturbations can be appreciated for the $x_{i2}$ state principally in MA1 and MA3. This variation correlates with the intermittencies depicted in Figure \ref{fig:NNnumerical} in the numerical simulation. Figure \ref{fig:circuit_G1x2} d) shows the projection of the systems in order to appreciate the attractor resulting from the network, $x_{11}$ vs $x_{22}$ is located in the upper part in green color, while $x_{11}$ vs $x_{32}$ is located in the lower part in orange color. Notice the resemblance between the projected synchronized attractors.

%

\begin{table}[!b]
\centering
\caption{ Values for the resistors of the coupling circuit from Figure \ref{fig:node1_coupling} for the two different types of coupling given in Figure \ref{fig:Grafos_circuito}.}
\label{tab:valores3}
\begin{tabular}{|c|c|c|c|}
\hline
\multicolumn{2}{|c|}{Nearest neighbor}                               & \multicolumn{2}{c|}{Fully connected}                                \\ \hline
\multicolumn{1}{|c|}{Component} & \multicolumn{1}{c|}{Value or name} & \multicolumn{1}{c|}{Component} & \multicolumn{1}{c|}{Value or name} \\ \hline
R18, R30, R41                   & 5k$\Omega$                         & R18. R30, R41                  & 3.3k$\Omega$                       \\ \hline
R10, R18, R29                   & 10k$\Omega$                        & R10, R18, R28, R29             & 10k$\Omega$                        \\ \hline
R24, R30, R36                   & 10k$\Omega$                        & R24, R30, R35, R36             & 10k$\Omega$                        \\ \hline
R27, R41, R47                   & 10k$\Omega$                        & R27, R41, R46, R47             & 10k$\Omega$                        \\ \hline
R28, R35, R46                   & Not connected                      &                                &                                    \\ \hline
\end{tabular}
\end{table}

\subsection{Electronic implementation of a fully connected network}

\begin{figure}
    \centering
    \begin{subfigure}[t]{0.47\textwidth} 
    \caption{}
        \includegraphics[width=\textwidth]{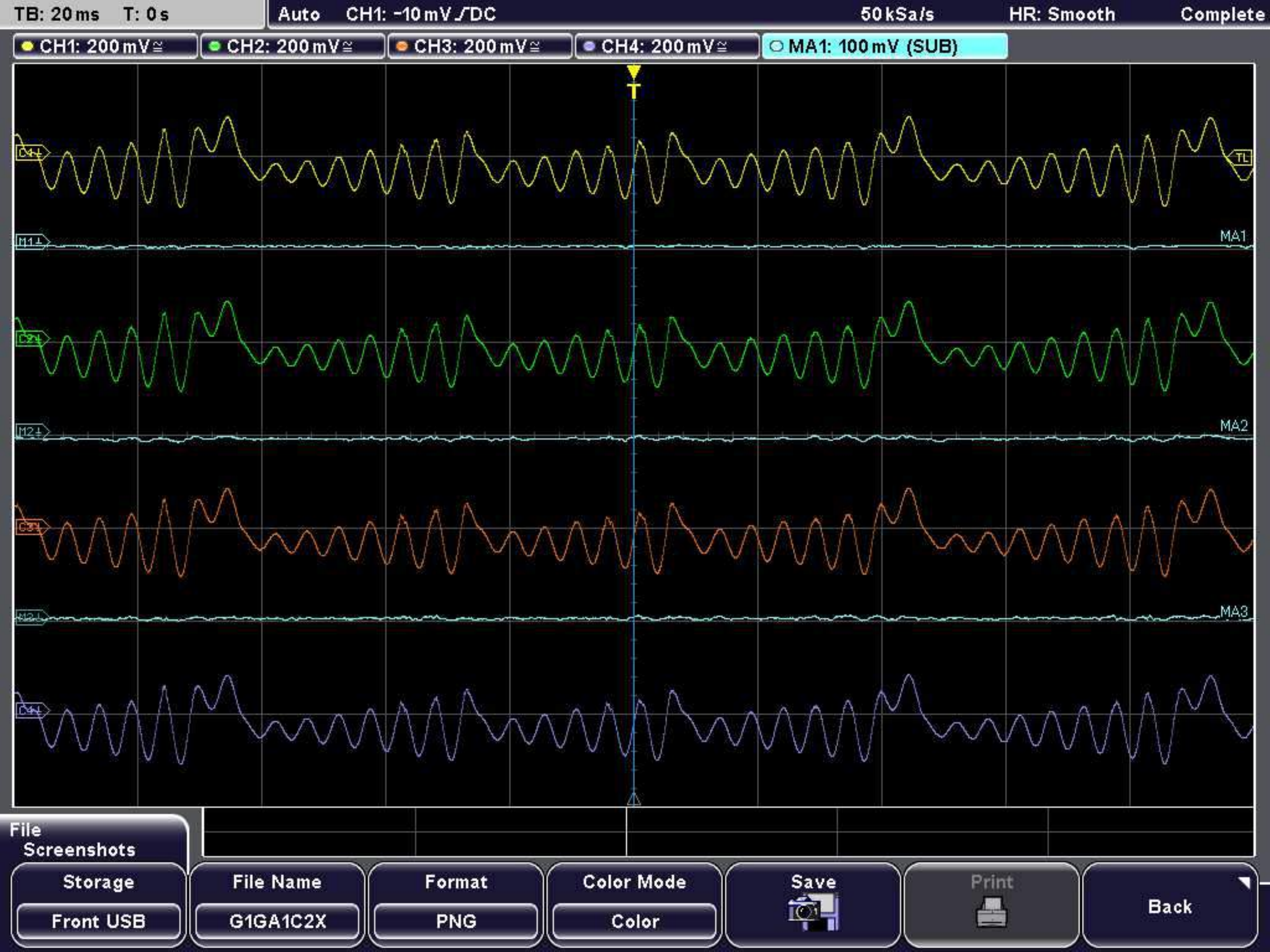}
    \end{subfigure}
    \begin{subfigure}[t]{0.47\textwidth} 
    \caption{}
         \includegraphics[width=\textwidth]{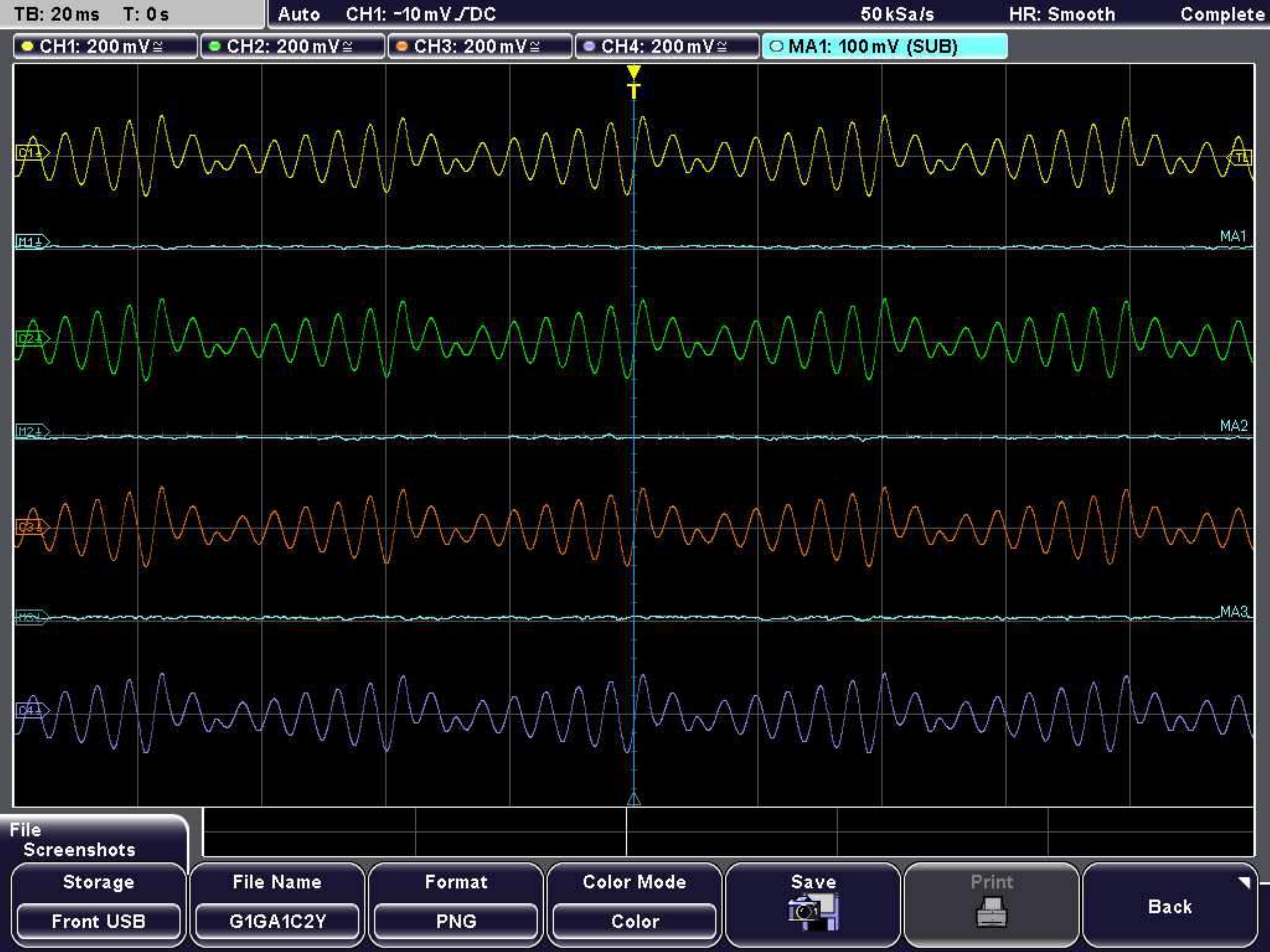}
    \end{subfigure}
        \begin{subfigure}[t]{0.47\textwidth} 
        \caption{}
        \includegraphics[width=\textwidth]{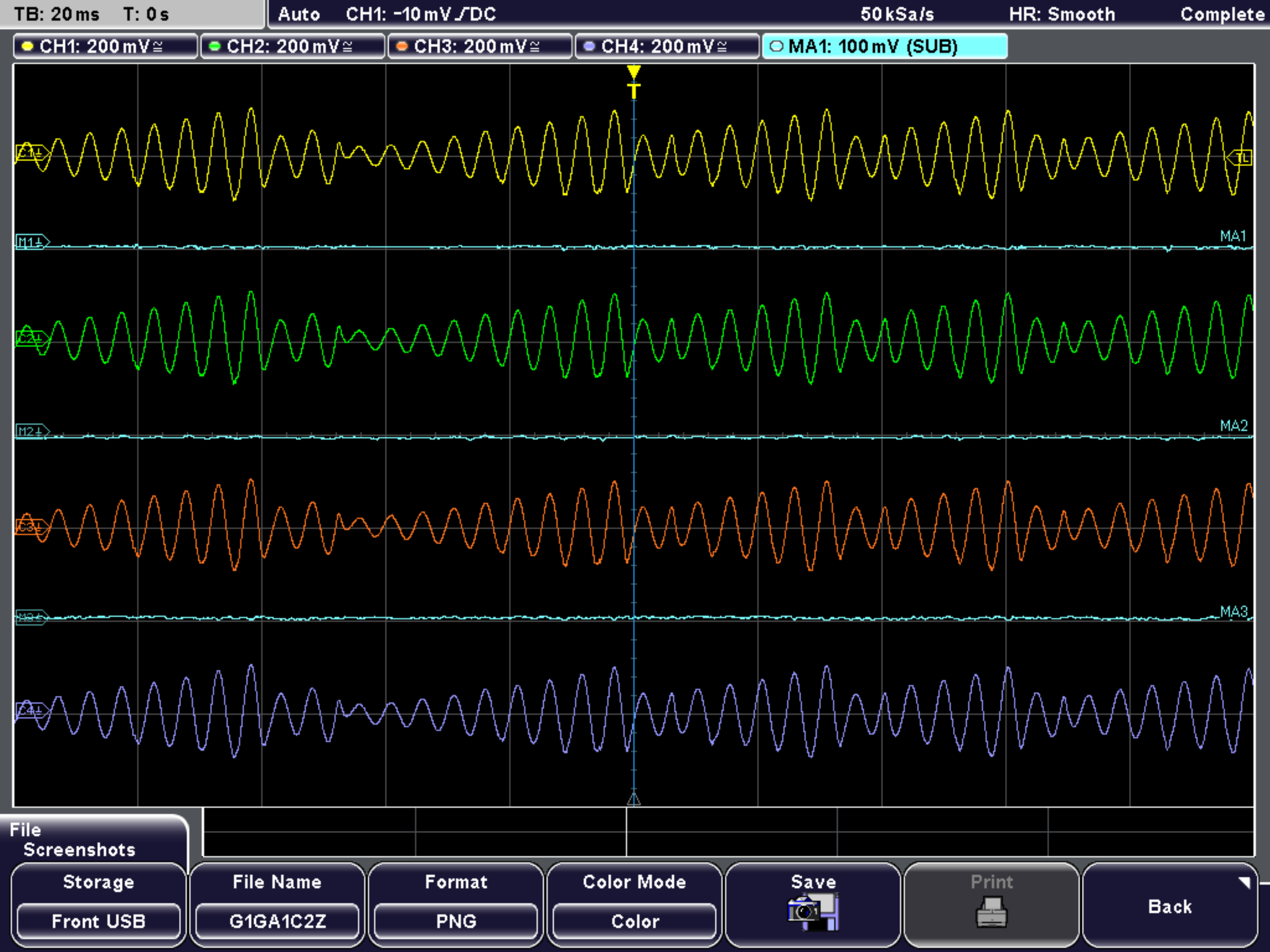}
    \end{subfigure}
    \begin{subfigure}[t]{0.47\textwidth} 
    \caption{}
         \includegraphics[width=\textwidth]{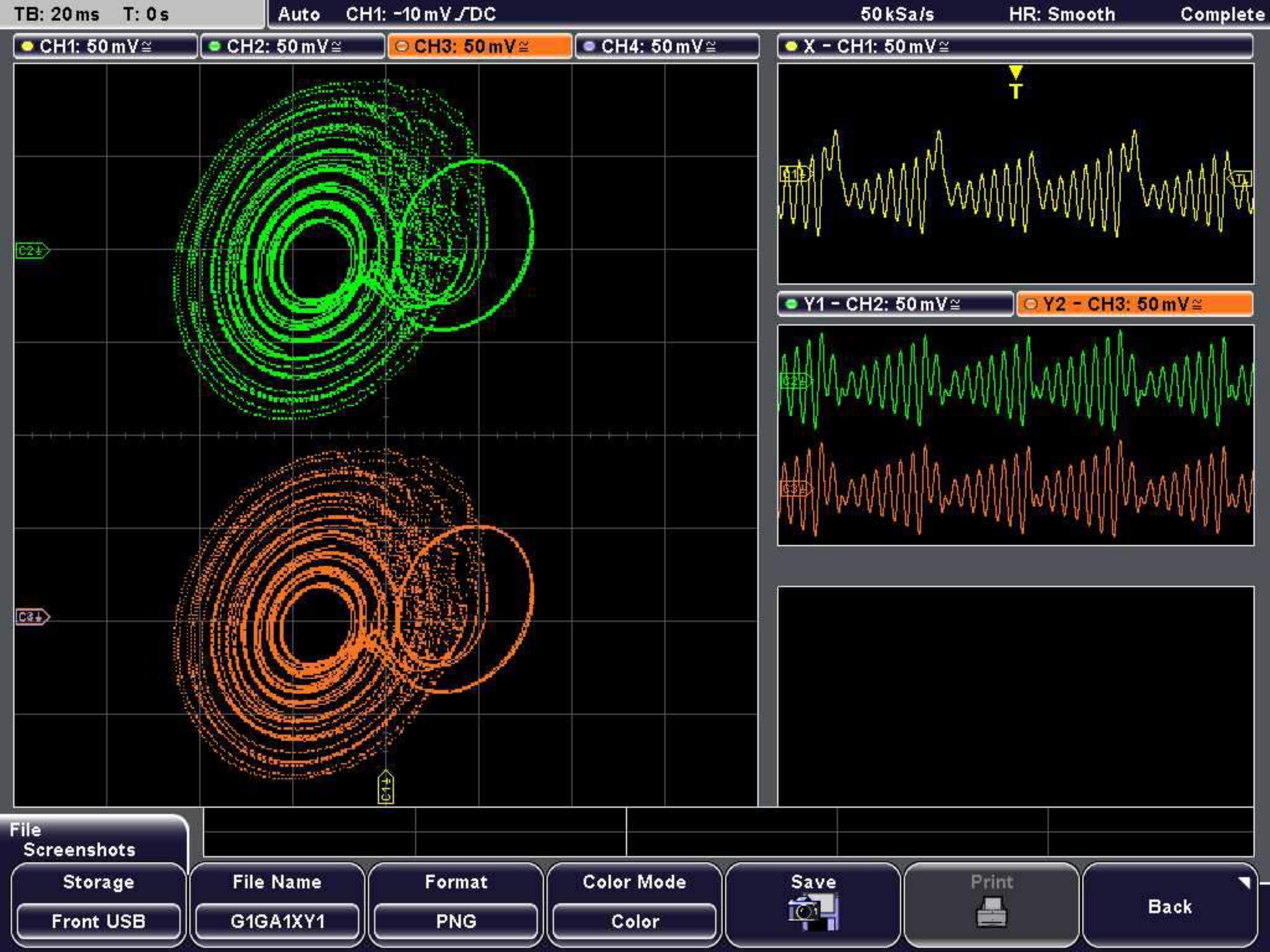}
    \end{subfigure}
\caption{Experimental traces from the oscilloscope for the signals of a) $x_{i1}$, b) $x_{i1}$, c) $x_{i1}$ of a $N=4$  fully connected neighbor network implemented electronically in the circuits depicted from Figures \ref{fig:node1_x1y1}-\ref{fig:node1_coupling}, with $\Delta^{fully}$ in eq. \eqref{eq:FCcircuito}, $\Gamma=diag{\mathbf{I}_3}$ with $c=2$. In d) is depicted the projection of the signals into the $(x_{11},x_{22})$ plane and the $(x_{11},x_{32})$ plane. The values for the resistor in the coupling of Figure \ref{fig:node1_coupling} are given in Tables \ref{tab:valores2} and \ref{tab:valores3}.}
\label{fig:circuit_G1x}
\end{figure}

By changing the resistors as the values depict in Table \ref{tab:valores3}, it will result in a fully connected network with a coupling matrix given by $\Delta^{fully}$ in eq. \eqref{eq:FCcircuito} with $N=4$, $\Gamma=diag{\mathbf{I}_3}$ and $c=2$. Figure \ref{fig:circuit_G1x} a) to d) depicts the experimental traces for each corresponding state $x_{i1},x_{i2},x_{i3}$. Same consideration according to the signals displayed here, from which it can be seen that the difference between the states has been attenuated in relation with the previous experiment. This experimental traces demonstrates the results in the numerical example of the fully connected network displayed before.
As mentioned before, devices commonly distributed present tolerances between 5-10\%. In thus,  the implementation of electronic devices with tolerance can be adequately adjusted in the designing of networks of identical or non-identical nodes resulting in a more natural systems.

%

\section{Concluding remarks}

In this work a network of nearly identical coupled chaotic oscillators is described.  In particular, a type of piece-wise linear (PWL) system called Unstable Dissipative System (UDS) was implemented, which is an affine-linear dynamical system with a switching law capable of {generating}  a chaotic behavior in the system. The main characteristic of this network is that it considered a tolerance variation in the parameters of each individual node, in order to present nearly identical states that resembles more accurately physical systems instead of identical simulated ones. The network was studied by means of numerical analysis and experimental validation by means of an electronic implementation using analog computing.

In this context, the model results in a network of nearly-identical nodes that, according to the numerical results,  achieve practical synchronous behavior. The circuit architecture let to select both a network topology and link attributes
and it was used to study experimentally the collective behavior of an ensemble of connected UDS, and to corroborate the emergence of bounded synchronization {which was defined as practical synchronous collective behavior}. Additionally, based on the formalism of dynamical networks, the  circuit dynamics were modeled by introducing parameter mismatches in the nodes, which emulate the natural tolerances in the nominal values of the circuit components. This type of electronic circuit device has potential applications in communications and  cryptography with the advantage of its relative easy implementation. Also it could be used to study the synchronization phenomena in systems of non-identical nodes by considering that each node has a distinct switching law \textit{i.e.}, different number of scrolls. These research issues will be reported elsewhere.

\section{Acknowledgements}
M. Garc\'ia-Mart\'inez acknowledges the GIEE - Optimizaci\'on y Ciencia de Datos for the support.
L.J.~Onta\~n\'on-Garc\'ia acknowledges the FAI-UASLP financial support through project No. C18-FAI-05-45.45 and the support given by the financing of the project UASLP-CA-268 with IDCA 28234  by SEP-PRODEP.

\section{Supplementary material}
An article brief description can be followed by means of the audio slides posted in https://youtu.be/VSuQCn7bdt4

\section*{References}

\end{document}